\documentclass{comjnl}

\usepackage{url}
\usepackage{enumerate}
\usepackage{graphicx}
\usepackage{epstopdf}
\usepackage{multirow}
\usepackage{subfigure}
\usepackage{amsmath,amssymb}
\usepackage{booktabs}
\usepackage{multirow}
\usepackage{makecell}
\usepackage{caption2}

\usepackage{graphicx}  
\usepackage{float}  
\usepackage{subfigure}  

\begin{document}

\title[Multi-View Dynamic Heterogeneous Information Network Embedding]{Multi-View Dynamic Heterogeneous Information Network Embedding}
\author{Zhenghao Zhang}
\affiliation{School of Computer Science and Technology, Xidian University, Xi’an, 710071, China} \email{zhangzhenghao1108@126.com}

\author{Jianbin Huang\textsuperscript{*}}
\affiliation{School of Computer Science and Technology, Xidian University, Xi’an, 710071, China} \email{jbhuang@xidian.edu.cn}
\shortauthors{Jianbin Huang}

\author{Qinglin Tan}
\affiliation{School of Computer Science and Technology, Xidian University, Xi’an, 710071, China} \email{zhangzhenghao1108@126.com}



\begin{abstract}
Most existing Heterogeneous Information Network (HIN) embedding methods focus on static environments while neglecting the evolving characteristic of real-world networks. Although several dynamic embedding methods have been proposed, they are merely designed for homogeneous networks and cannot be directly applied in heterogeneous environment. To tackle above challenges, we propose a novel framework for incorporating temporal information into HIN embedding, denoted as Multi-View Dynamic HIN Embedding (MDHNE), which can efficiently preserve evolution patterns of implicit relationships from different views in updating node representations over time. We first transform HIN to a series of homogeneous networks corresponding to different views. Then our proposed MDHNE applies Recurrent Neural Network (RNN) to incorporate evolving pattern of complex network structure and semantic relationships between nodes into latent embedding spaces, and thus the node representations from multiple views can be learned and updated when HIN evolves over time. Moreover, we come up with an attention based fusion mechanism, which can automatically infer weights of latent representations corresponding to different views by minimizing the objective function specific for different mining tasks. Extensive experiments clearly demonstrate that our MDHNE model outperforms state-of-the-art baselines on three real-world dynamic datasets for different network mining tasks.
\end{abstract}

\maketitle

\section{Introduction}
\label{section1}
In recent years, network embedding has attracted increasing research attention and been proved extremely useful in analyzing and mining networks. Network embedding aims at seeking proper mapping functions so that original high-dimensional sparse network can be embedded into low-dimensional vector space where the proximities among nodes are preserved. Under the circumstance that all nodes and links of different types can be represented as low-dimensional dense vectors, not only is memory space greatly saved, but such low dimensional representations can directly serve as the feature inputs of machine learning algorithm and help to efficiently enhance the performance of various mining tasks. This brings great convenience to network analysis. 

Although researches on learning representations of Heterogeneous Information Network (HIN) have progressed rapidly in recent years, most of these existing HIN embedding methods are predominately designed for static environments and neglect the temporal information(e.g., evolving patterns and dynamic interactions) in network data.

As the saying goes, nothing is permanent but change. Real-world HINs are ubiquitous in domains, ranging from social networks (e.g., WeChat, Facebook), scientific collaboration networks (such as DBLP), to biological network(e.g., protein-protein interaction network), which feature complicated topological structure and rich semantic relationships, together with adding/deleting of links and nodes of different types(\cite{zhuang2013influence},\cite{li2016toward}). For instance, in social network, new users can be added into existing network, and establish a new friendship with existing users, moreover, old friendships may be deleted over time. Compared to static networks, dynamic networks are more precise in characterizing the complex and dynamic systems. Thus, the analysis of dynamic network has attracted considerable attention(\cite{zhou2018dynamic},\cite{tong2008colibri},
\cite{tang2008community},\cite{ning2007incremental}
\cite{aggarwal2011node}).

However, as far as we have known, no dynamic HIN embedding algorithm has been proposed so far. Considering the new challenges that continuously added or removed nodes and links of different types bring to HIN embedding, existing dynamic embedding methods, including Dynamic-Triad \cite{zhou2018dynamic} and Dyn-GEM  \cite{goyal2018dyngem}, which are merely designed for homogeneous network, can$'$t be directly applied in heterogeneous environments. 
Naively and independently learning representations for each snapshot with static embedding algorithms, and then rotationally conducting the learned representations across time steps(\cite{zhu2016scalable},\cite{zhang2018timers}), will lead to undesirable performance. Computational complexity of learning representations for every single snapshot repetitively is also very high. More importantly, these approaches can$'$t capture the evolving patterns and interaction between nodes across the time steps. The evolutionary patterns provide novel insights into preserving the structure and semantic relationships of dynamic HIN, and how to update different types of representations and temporally preserve rich semantic information into latent vector space turns out to be problem worth investigating. 

Moreover, in HINs, especially on dynamic environment, semantic relationship instances from a specific view reflected by meta-path are biased due to sparsity. Therefore, proximities from diverse views are demonstrated to be important in capturing dynamic interactions internal and across the time steps. However, it is much more challenging for network embedding methods to efficiently incorporate the newly added/deleted edges, because any changed links will affect the proximities between nodes guided by different meta-paths. How to select and fuse the semantic proximity from different views is also an open problem. It is required to design a novel semantic based proximity measure which can discover the subtle differences of neighbors and learn their relation strength accurately.

To tackle the aforementioned challenges, we propose an efficient and stable embedding framework for dynamic HIN, referred to as Multi-view Dynamic Heterogeneous Network Embedding (MDHNE), which serves as a basis to incorporate temporal dependencies from multiple views into existing HIN embedding based on Recurrent Neural Network (RNN) and the attention mechanism, and thus temporally derive the updated node representations while preserving the proximities of different views.

Our proposed MDHNE extends deep RNN model into a sparse dynamic heterogeneous information network scenario. We apply the deep recurrent architecture to capture highly complex and dynamic temporal features. After a series of non-linear functions in the recurrent layers of RNN respectively, transformation patterns of structure and dynamically changed proximities can be embedded in latent vector space, and thus the node representations of multiple views can be updated over time. Moreover, we come up with an attention based deep fusion mechanism which can automatically infer weights of latent representations corresponding to different views according to the objective function of specific data mining task. The whole model can be efficiently trained through the back-propagation algorithm, alternating between optimizing the view-specific node representations and voting for the robust node representations by learning the weights of different views. 

To verify the advantages of our proposed algorithm, we conduct extensive experiments on three real-world dynamic datasets. As indicated by experimental results, our proposed approach significantly outperforms other representative embedding methods in various network mining tasks such as node classification and recommendation task, which means that our proposed method is able to accurately learn and update the representations of vertices with network evolves and preserve the proximities affected by changed links well from dynamic HINs.

The major contributions of our work can be summarized as follows:
\begin{itemize}
	\item To our best knowledge, this is the first attempt to study the dynamic heterogeneous information network embedding. Our proposed model uses deep  RNN encoder to incorporate temporal transforming patterns of structure and interactions between nodes from different views into latent vector spaces, and thus the node representations from multiple views can be learned and updated over time.
	\item We propose an attention based multi-view fusing mechanism, which can automatically infer the weights of latent representations corresponding to different views and vote for the final node representations more comprehensive and accurate.
	\item We conduct extensive experiments on various real-world HINs. Experimental results on two tasks prove the effectiveness and efficiency of our proposed approach over many competitive baselines. 
\end{itemize}

The rest of this paper is organized as follows. Section \ref{sec:related-works} briefly reviews related work. Then preliminary and some related definitions are introduced in Section \ref{sec:proposed-model}. In Section \ref{sec:Multi-View Dynamic HIN}, the detailed descriptions of multi-view dynamic heterogeneous network are given. Then, a novel dynamic heterogeneous network embedding approach referred to as MDHNE is presented in Section\ref{sec:Multi-View Dynamic HIN Embedding}. In Section \ref{sec:Experimental evaluation}, dataset descriptions and experimental evaluations are reported. Conclusion and future research direction will be presented in Section 7.

\section{Related work}
\label{sec:related-works}
Network embedding, i.e., network representation learning (NRL), is proposed to embed network into a low dimensional space while preserving the network structure and property so that the learned representations can be applied to the downstream network tasks. We will introduce the progress of HIN embedding and dynamic network embedding respectively. 

\subsection{Heterogeneous Information Network Embedding}\label{subsecHINModel}
In present, network embedding methods are primarily divided into two categories according to network types. One is homogeneous network embedding, and the other is heterogeneous network embedding. 
Homogeneous network embedding mainly consist of random walk based methods(\cite{perozzi2014deepwalk},\cite{grover2016node2vec},\cite{cao2015grarep}), deep neural network based methods\cite{wang2016structural},\cite{tang2015line}, and matrix factorization based methods\cite{belkin2003laplacian},\cite{xue2017deep}. In a homogeneous information network, there exists only one single type of nodes and the nodes can walk along any path. Comparatively, heterogeneous information network embedding which is seldom studied before has attracted growing interests in recent years. Metapath2vec\cite{dong2017metapath2vec} implements meta-path guided random walk method and utilizes both the skip-gram algorithm and negative sampling to learn heterogeneous network node embedding. HINE\cite{shi2018heterogeneous} firstly calculates proximities between nodes by means of meta-path based random walk and then adopt the proximity information as supervisory information to learn node representations.  Unfortunately, the above method can only capture relatively simple and biased semantic information of nodes, so a new neural network based HIN embedding model known as HIN2Vec\cite{fu2017hin2vec} has been raised which takes nodes of different types and complex diversified relations among nodes into account. HIN2Vec respectively learns node and meta-path latent representations by predicting relations among nodes. Esim\cite{shang2016meta} attempts to capture various semantic relations of nodes through multiple meta-paths. Moreover, Zhang comes up with Metagraph2vec\cite{zhang2018metagraph2vec}, trying to use meta-graphs as guidelines for random walk. Richer structural details and more complete semantic information are successfully extracted. Besides, there are also some other HIN embedding methods designed for some particular tasks, such as identifying authors\cite{chen2017task} , recommendation\cite{wang2018shine},\cite{tang2015pte}. However, all the aforementioned approaches only focus on static HIN embedding.

\subsection{Dynamic Network Embedding}\label{subsecHINModel}

From the view of network status, embedding techniques can be broadly divided into two catagories: 1.static network embedding, which embed each node into a unified latent vector space; 2. dynamic network embedding, which considers multiple snapshots of a graph and obtains a time series of vectors for each node. 
Most analyses have been done on static embedding and dynamic embedding as a new topic is still under investigation. Several methods have been proposed to extend static graph embedding approaches by adding regularization\cite{zhu2016scalable}. Over the past several years, some approaches have been proposed to study dynamic graph embedding. Specifically, Zhu et al.\cite{zhang2018timers} developed a dynamic network embedding algorithm based on matrix factorization. DynamicTriad\cite{zhou2018dynamic} models the triadic closure process to capture dynamics and learns node embedding at each time step, and it relaxes the temporal smoothness assumption but only considers the spanning of two steps. In addition, DynGEM\cite{goyal2018dyngem} use the learned embedding from the previous time step graphs to initialize the current time step embedding. Although it does not explicitly use regularization, such initialization implicitly keeps the embedding close to the previous. Song et al. extend skip-gram based models and propose a stable dynamic network embedding framework with high efficiency. Recently, Rakshit\cite{trivedi2018representation} uses a warm start method to train across snapshots and employs a heuristic approach to learn stable embeddings over time but do not model time. DANE\cite{li2017attributed} proposes an embedding method based on perturbation theory in dynamic environment with nodes' attributes changing over time. Besides, there are also some task-specific temporal network embedding methods. NetWalk\cite{yu2018netwalk} is an anomaly detection framework, which detects network deviations based on a dynamic clustering algorithm. 

All above-mentioned dynamic approaches are designed to handle homogeneous networks. None of them integrate both of heterogeneity and dynamics into network embedding. The dynamic changes of heterogeneous networks are much more complex than homogeneous networks. We are eager to design embedding methods specific for the characteristics of dynamic HINs. 
                  
\section{Preliminaries}\label{sec:proposed-model}
We first introduce some relevant definitions used in this paper and give the brief description of dynamic HIN embedding.

\subsection{Dynamic Heterogeneous Information Network}
\label{sec:encoder-decoder}

Real-world HINs are dynamic in nature and evolve over time, together with increasing/decreasing/changing of links and nodes of different types. Movielens is taken as an example to illustrate the network which is both dynamic and heterogeneous. When a new node denoted as a new movie joins the network, corresponding new nodes denoted as new actor nodes, director nodes, and tag nodes also join the network. Meanwhile, new links will be built between these new nodes of different types, and these links represent the acted, directed and containing relationship. Moreover, these links may be connected of the existing nodes or the new nodes. 

Next, we will define the dynamic heterogeneous network in detail: given a temporal heterogeneous information network $G = (V,E,A,R)$, it can be slice into a series of HIN snapshot denoted as $G{\rm{ = \{ }}{G_{\rm{1}}}{\rm{,}}{G_2}...{\rm{,}}{G_T}{\rm{\} }}$, which represents the state of the network from time step 1 to $T$ and $T$ denotes the time window size. Therefore, we represent the temporal HIN at the time step $t$ as 
${G_t} = ({V_t},{E_t},A,R)$, which consists of the set of nodes
${V_t}$, the set of links ${E_t}$ between nodes, the set of node types $A$ and the set of link types $R$ respectively. In usual cases, it is associated with a node type mapping function 
$\Phi :{V_t} \to A$ which represents that each vertex $v \in {V_t}$ can be mapped into a node type $a \in A$, and a link type mapping function is expressed as $\Psi :{E_t} \to R$, meaning that each link $e \in {E_t}$ can be mapped into an edge type
$r \in R$.
In this paper, for the ease of presentation, we follow the assumption that nodes of various types remain constant and links change when network evolves.

\subsection{Dynamic HIN Embedding}
\label{sec:encoder-decoder}

Given an evolution of HIN $G{\rm{ = \{ }}{G_{\rm{1}}}{\rm{,}}{G_2}...{\rm{,}}{G_T}{\rm{\} }}$ , we aim to learning a time series of mapping functions $\Omega  = ({\Omega _1},{\Omega _2},...,{\Omega _T})$  to embed each node of different types into low-dimension embedding space, so that the learned representations at each time step have the capability to effectively capture rich semantic relationships among nodes and preserve the non-linear historical evolution information. Specifically, for temporal HIN ${G_t} = ({V_t},{E_t})$  at any time step t, by learning a mapping function ${\Omega _t}:{V_t} \to {R^d}$ , each node $v \in {V_t}$ can be represented into a latent d-dimension representation ${x_t} \in {R^d}$ with $d \ll |V|$ . Therefore, we can continuously update the old embedding results and keep the accuracy of HIN embedding. 

\subsection{Semantic Structure of HIN}
\label{sec:encoder-decoder}

Complexity of the heterogeneous network forces us to propose a structure, which can describe meta-level semantic information of a heterogeneous information network, so that node types and relations between nodes in the network can be better comprehended. 

Definition 1 (Network Schema) Given a heterogeneous information network, a network schema can be abstracted which is referred to as ${\Theta _G} = (A,R)$ . Where ${\Theta _G}$  is a directed graph that contains all allowable node and link types, these types are combined together based on a given schema template to conduct meta-level semantic description of the network $G$. Moreover, both meta-path and meta-graph are semantic structures generated from HIN schema. 

\begin{figure}[ht]
	\centering
	\subfigure[DBLP-schema]{
		\begin{minipage}[t]{0.5\linewidth}
			\centering
			\includegraphics[width=1\linewidth]{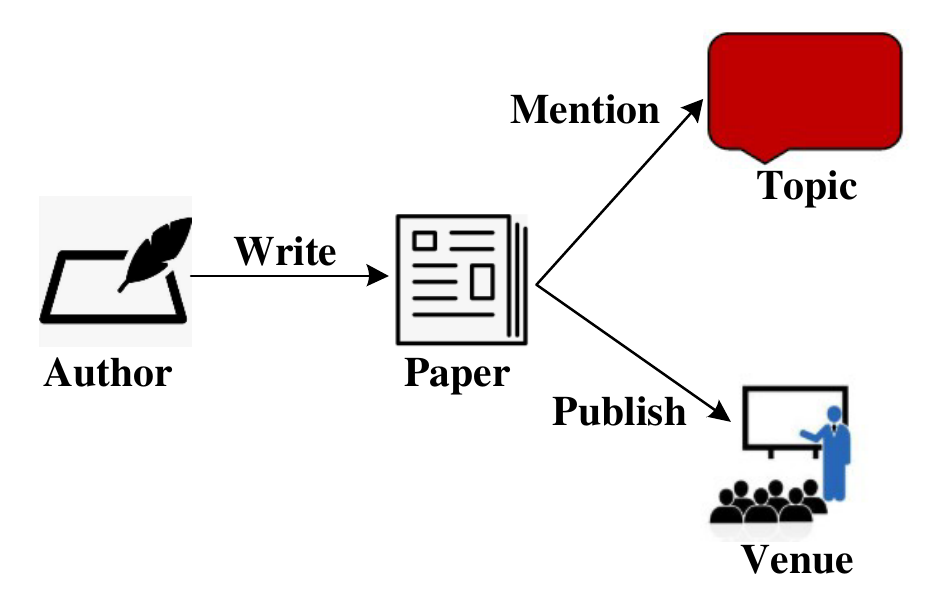}
		\end{minipage}%
	}%
	\subfigure[movielens-schema]{
		\begin{minipage}[t]{0.5\linewidth}
			\centering
			\includegraphics[width=1\linewidth]{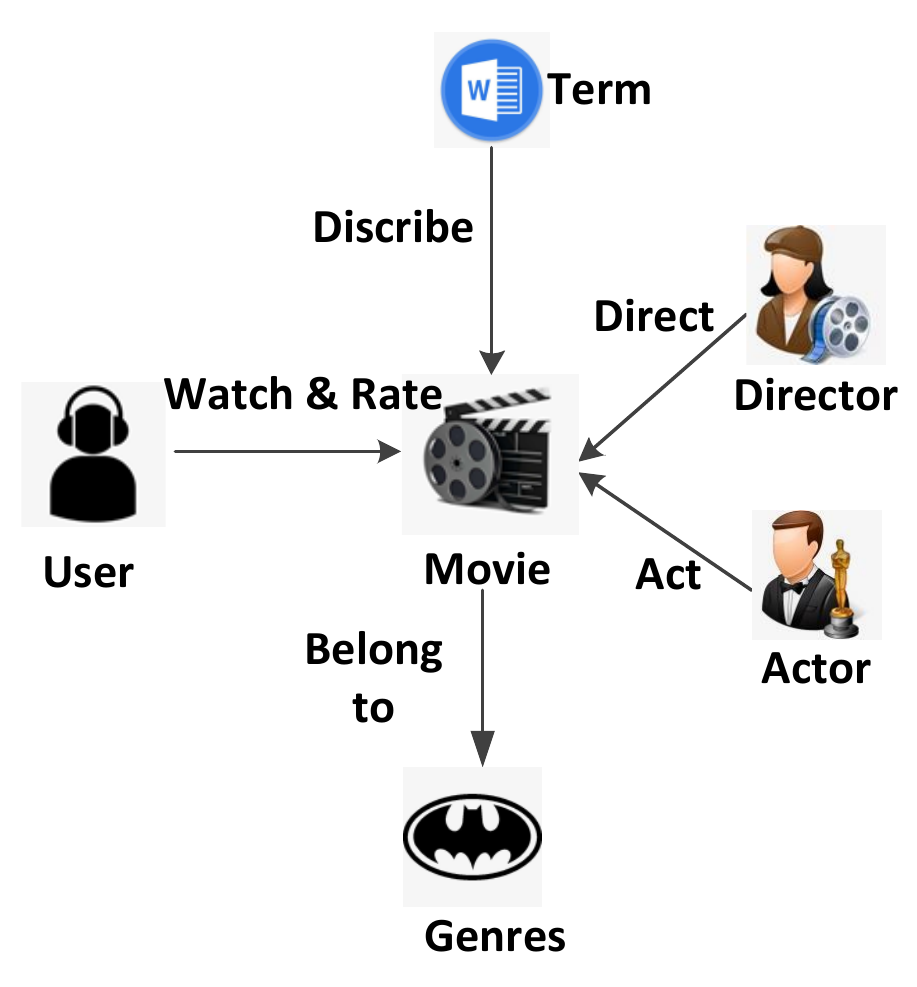}
		\end{minipage}%
	}%
	\centering
	\caption{Network schema of different  HIN datasets.}
\end{figure}

To be more specific, in Fig.1(a), DBLP HIN schema consists of 4 node types (A for authors, P for papers, V for venue and T for keywords), and 3 link types (an author-paper link signifying papers published by the author. A paper- venue link signifying a venue on which a paper is published. A paper-keyword link signifying keywords of papers). By contrast, in Fig.1(b), MovieLens HIN schema comprises 6 node types of U (users), M (movies), A (actors), D (directors), T (tags) and G (genres) and 5 link types including users watching and reviewing movies, actors in movies, directors directing movies, tags of movies and cinematic genres.

Definition 2 (Meta-path) As an abstract sequence of node types connected by link types, the meta-path is formed by transforms of a network schema and able to capture rich semantic information preserved in heterogeneous information networks. Specifically, given a HIN schema denoted as ${\Theta _G} = (A,R)$ , a meta-path can be expressed in the form of 

\begin{equation}\label{MD}
P = {a_1}\mathop  \to \limits^{{R_{1,2}}} {a_2}\mathop  \to \limits^{{R_{2,3}}} ...{a_{l - 1}}\mathop  \to \limits^{{R_{l - 1,l}}} {a_l},
\end{equation}
where ${a_i} \in A(i = 1,...,l)$ indicates node types and  ${r_{i,j}} \in R$ represents link types between ${a_i}$ and ${a_j}$, $1 \le i,j \le l$.

For example in Fig.2, the meta-path U-M-G-M-U in the movie review network MovieLens indicates that movies rated by two users contain the same genres. In addition, U-M-A-M-U and U-M-D-M-U respectively mean that movies rated by two users are acted by the same actor and have the common director. Clearly, different meta-paths represent different semantics. 

\begin{figure}
	\centering
	\includegraphics[width=\linewidth]{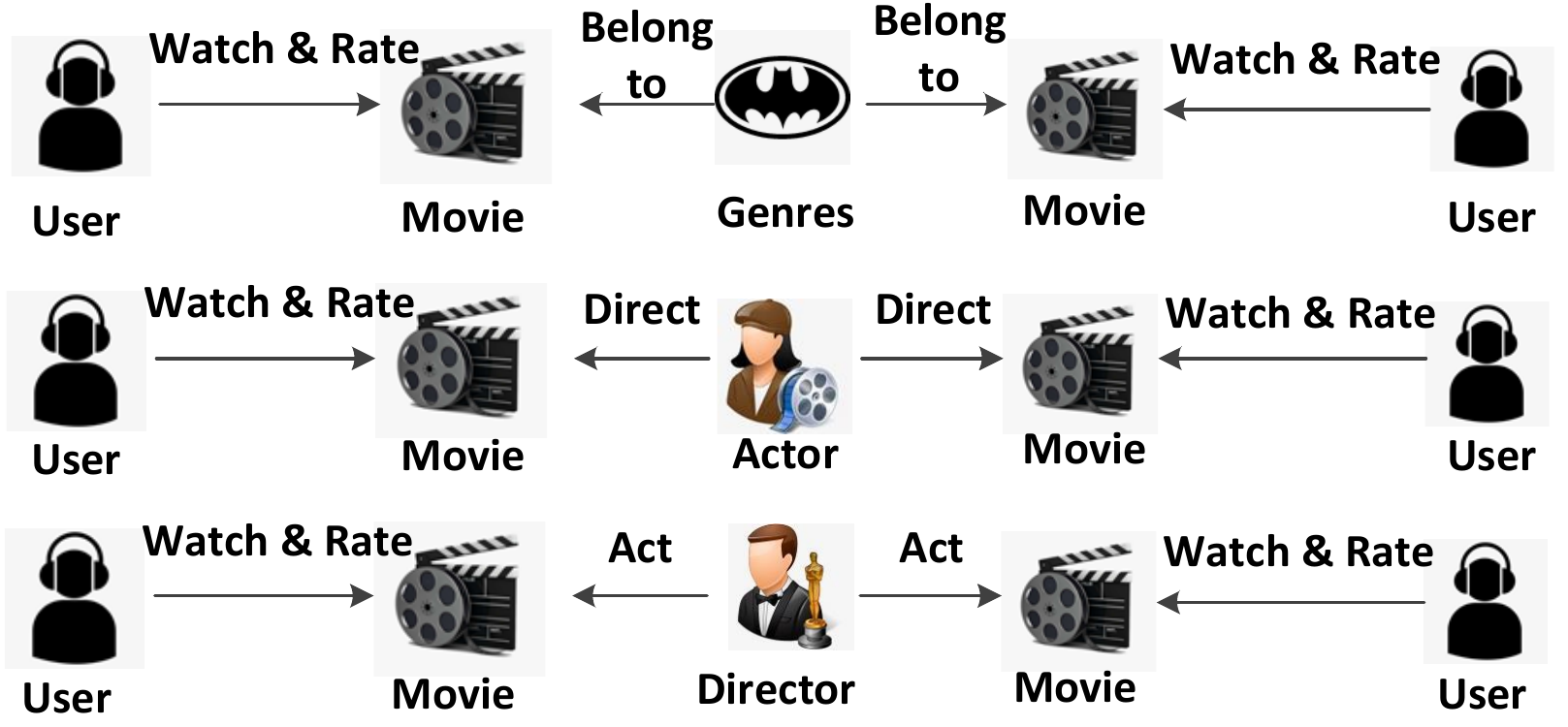}
	\caption{Meta-path examples of MovieLens.}
	\label{fig:datasets}
\end{figure}

In this case, the meta-path has an ability to learn various types of important semantic information in the network and can be applied in heterogeneous network mining tasks. 

\section{Multi-View Dynamic HIN}\label{sec:Multi-View Dynamic HIN}

In reality, it is observed that heterogeneous information network contains different types of semantic relationships and each type of semantic relationship defines a network view. However, most of existing HIN embedding models merely focus on one single semantic relationship which is sparse and biased to some extent. In order to make the learned node representations more accurate and comprehensive, especially on dynamic environments, semantic relationships from different views in HIN must be fully taken into account when constructing the network embedding model. Therefore, in this paper, a series of meta-paths ${\Phi _0},{\Phi _1},...,{\Phi _{|P|}}$ are selected to extract complex and diverse semantic relationships, and each meta-path corresponds to a specific network view. 

With a viewpoint of "Making Hard Things Simple", we choose to transform the original HIN into a series of homogeneous networks 
$\{ {G_{{\Phi _0}}},{G_{{\Phi _1}}},...,{G_{{\Phi _{|P|}}}}\} $ corresponding to a series of network views $\{ vie{w_1},vie{w_2},...,vie{w_{|P|}}\} $ respectively. For any homogeneous network, denoted as ${G_{{\Phi _p}}},(1 \le p \le P)$ , which corresponds to the specific $vie{w_p},(1 \le p \le P)$  and means that nodes in ${G_{{\Phi _p}}}$ are connected to each other through the specific meta-path ${\Phi _p}$ .

For instance, in the movie review network MovieLens, the meta-path U-M-U describes that two users rate a common movie, which corresponds to "co-rate" view. While another meta-path U-M-G-M-U means that movies rated by two users share the same genres, which can correspond to the "same genres" view. Moreover, meta-path U-M-A-M-U means that movies rated by two users have the same actors, which can correspond to the "common actors" view. Clearly, different meta-paths correspond to different semantic views. 

Based on the illustration above, it can be expanded to the dynamic and heterogeneous environments: As shown in Fig.3, given a series of HIN snapshots ${\rm{\{ }}{{\rm{G}}_1}{\rm{,}}{{\rm{G}}_2}{\rm{,}}...{\rm{,}}{{\rm{G}}_T}{\rm{\} }}$, which represent the state of the network at each time step and T denotes the time window size. Given the 
${G_t} = ({V_t},{E_t},A,R)$ , which represents the state of network at time step t, following the introduction in the previous paragraph, we transform ${G_t}$  into a homogeneous network set, denoted as $G_{{\Phi _0}}^t,G_{{\Phi _1}}^t,...,G_{{\Phi _{|P|}}}^t$ , which correspond to a series of network views $\{ vie{w_1},vie{w_2},...,vie{w_{|P|}}\} $ respectively. 

\begin{figure}
	\centering
	\includegraphics[width=\linewidth]{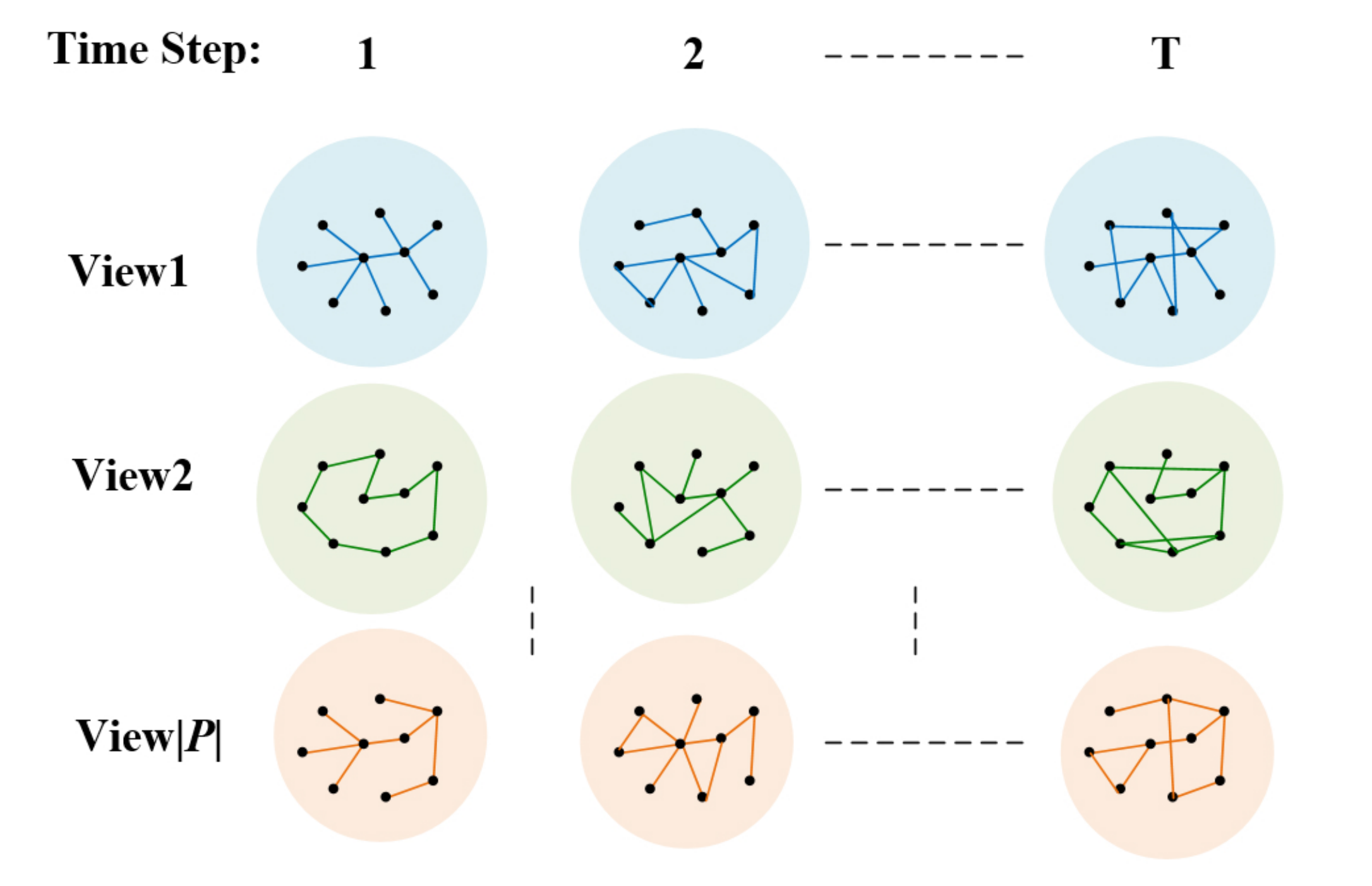}
	\caption{An example of multi-view dynamic HIN. Each view corresponds to a type of meta-path based connection between nodes, which is characterized by a set of links. Different views are complementary to each other.}
	\label{fig:datasets}
\end{figure}

Definition 4 (Meta-path Based Time-aware Commuting Matrix) Given the corresponding meta-path ${\Phi _p} = ({a_1},{a_2},...,{a_{l - 1}},{a_l})$ , we can define the meta-path based time-aware commuting matrix $M_{{\Phi _p}}^t$  at time step t as follows: 

\begin{equation}\label{MD}
M_{{\Phi _p}}^t = W_{{a_1},{a_2}}^t \times W_{{a_2},{a_3}}^t \times ... \times W_{{a_{l - 1}},{a_l}}^t,
\end{equation}
where $W_{{a_i},{a_{i + 1}}}^t$  represents the adjacent matrix between nodes of type ${a_i}$  and ${a_{i{\rm{ + }}1}}$  at the time step $t$. For source node ${v_1}$  of type  ${a_1}$ and target node ${v_l}$  of type ${a_l}$ , the element $M_{{\Phi _p}}^t(1,l)$ means the number of path instances guided by meta-path ${\Phi _p}$  from ${v_1}$  to ${v_l}$  at the time step $t$. 

Due to the dynamic nature of the network, links between different types of nodes add or delete over time, so the number of path instances between node pairs guided by meta-paths varies over time. 
In this paper, we focus on the problem of dynamic HIN embedding with multiple views proximities preserved. After transforming the temporal HIN into a series of dynamic homogeneous networks of different views, we aim to expand traditional RNN into a temporal HIN scenario to dynamically embed transformation patterns into latent representations of different views at each time steps. 

\section{MDHNE: Multi-View Dynamic HIN Embedding}\label{sec:Multi-View Dynamic HIN Embedding}

A successful dynamic network embedding method could employ graph evolution information to capture network dynamics and learn stable embedding over time. To capture the temporally evolving information about the nonlinear dynamics governing the changes in patterns of HIN topology and interactions between nodes, we develop a novel Dynamic multi-view HIN embedding framework (MDHNE). The proposed model can dynamically learn and update node representations from different views via deep temporal RNN encoder and effectively integrate them with an attention mechanism.

\begin{figure}
	\centering
	\includegraphics[width=\linewidth]{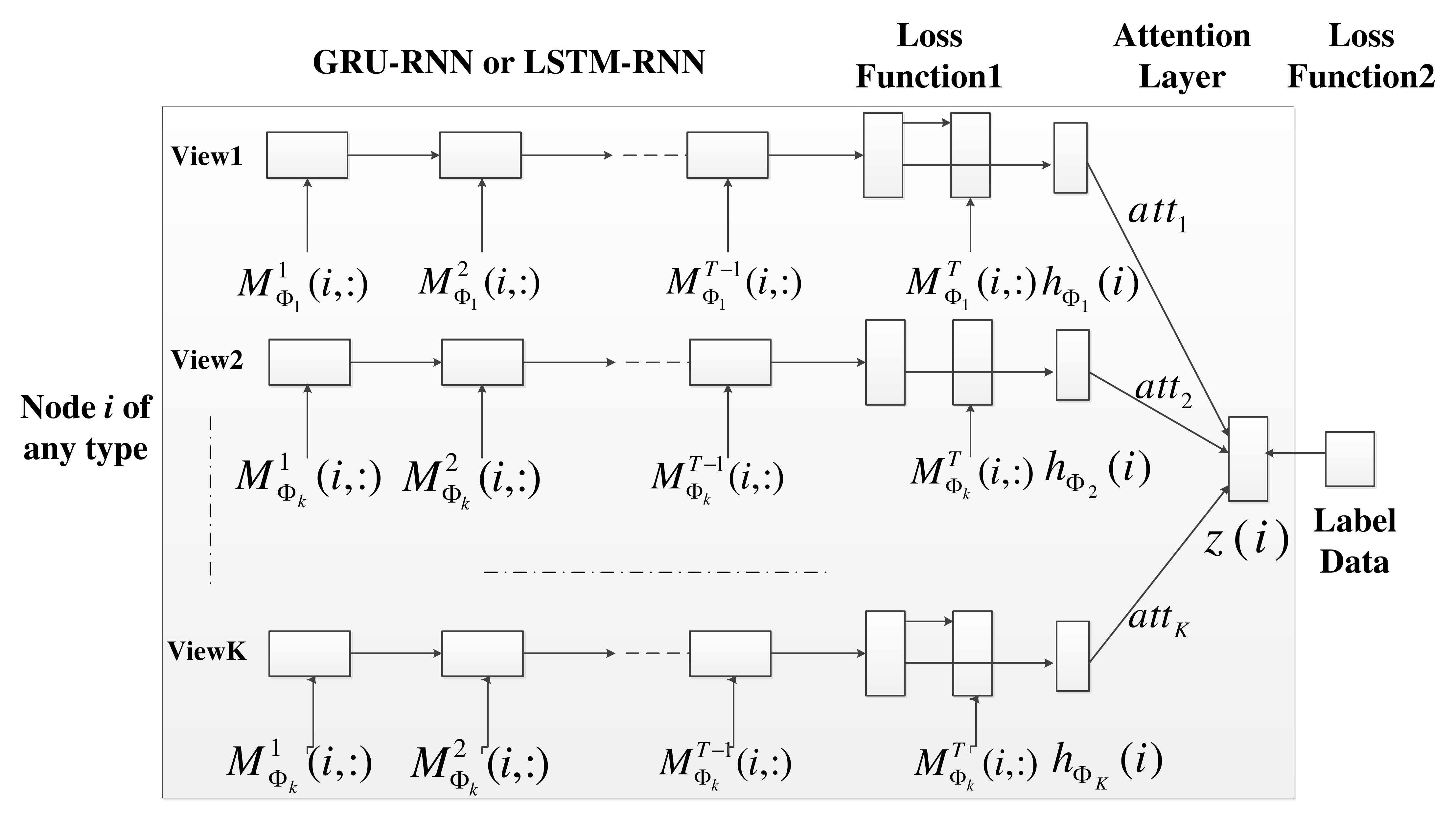}
	\caption{MDHNE framework for node classification task.}
	\label{fig:datasets}
\end{figure}

\begin{figure}
	\centering
	\includegraphics[width=\linewidth]{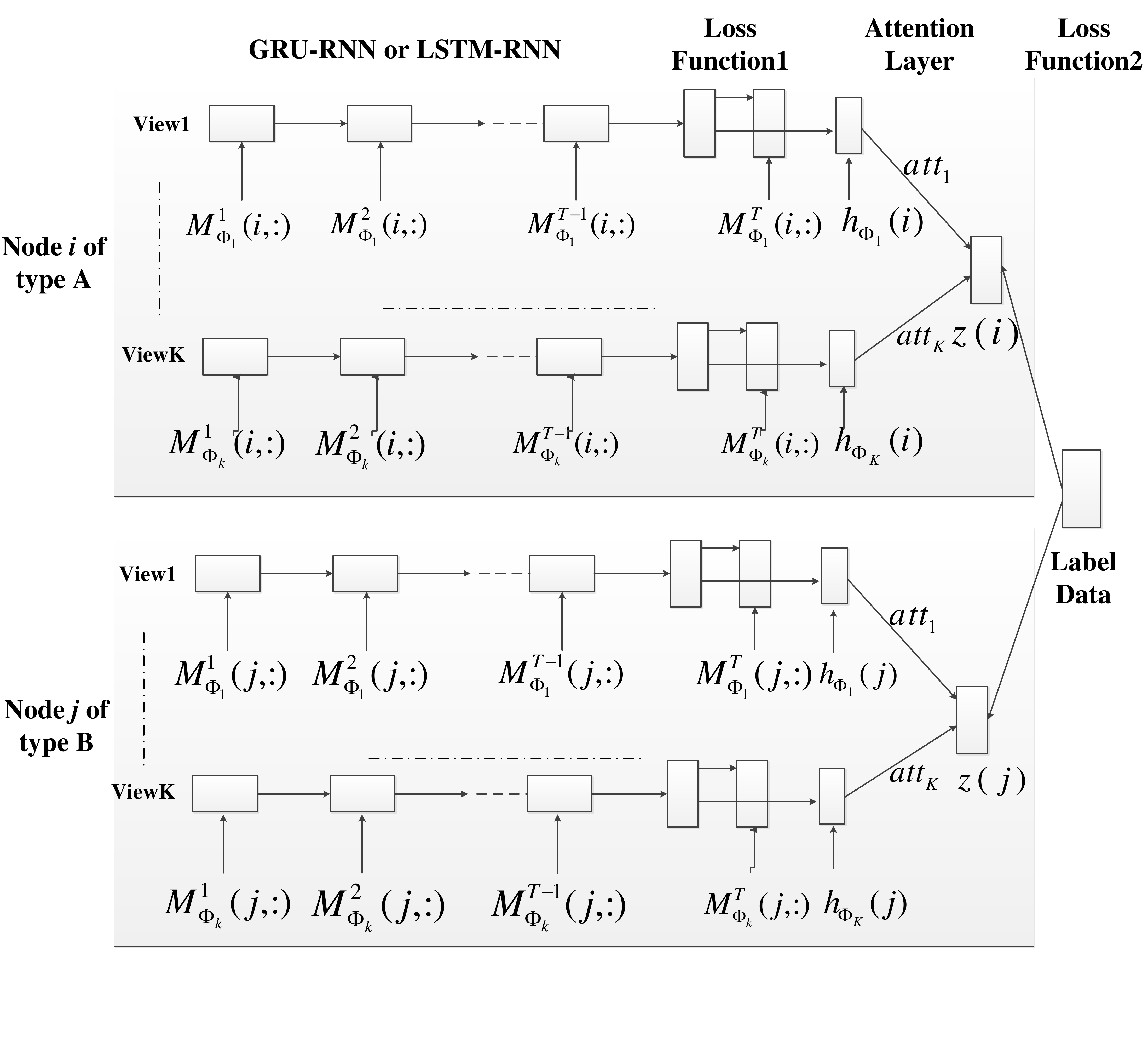}
	\caption{MDHNE framework for recommendation task.}
	\label{fig:datasets}
\end{figure}

The overall architecture of MDHNE model is principally described in Fig.4 and Fig.5, which mainly consists of three modules, i.e., dynamic latent vector learning module, embedding fusion module, and optimization module: 
\begin{itemize}
	\item In dynamic latent vector learning module, our MDHNE extends the traditional RNN into dynamic and sparse HIN. With regard to temporal similarity matrix of homogeneous network derived from different views in HIN at each time slice, we use them as the original input of each RNN unit at each time step to capture the high-order non-linear proximity information within and across time steps, and the latent node representations in different views can be updated over time.

	\item In fusion module, in order to integrate different views of latent representations for improving embedding quality, we utilize the attention mechanism to automatically infer the weights of the learned latent vectors during fusion, which will leverage a few labeled data (practical links).
	
	\item In optimization module, we minimize the loss function specific for different tasks to optimize the parameters in our model, so as to embed the evolving pattern of network structure and changed semantic relationships into node representations comprehensively and accurately. 
	
\end{itemize}

In this section, we briefly define the problem statement. And in the next section, multiple variations of MDHNE model capable of capturing temporal patterns in dynamic HIN will be introduced in detail, including RNN based temporal encoding process, fusing process via attention mechanism. Finally, loss functions and optimization approach will be further discussed. 

\subsection{RNN Based Temporal Encoder}\label{subsecSimilaritymeasure}
To solve the problem of feature extraction in dynamic network, we need to pay attention to the evolutionary history of the network, so that the learned latent representations reflect the changes made in the network over time. Traditional solution is to naively and independently learn representations for each snapshot with static embedding algorithms, and then stack all the learned representations from a series of time steps into one single representation. However, this approach will result in difficulties in the learning procedure due to the curse of dimensionality and lack the ability to capture the temporal dependencies across time steps. 

In dynamic networks, there can be long-term dependencies which may not be captured by traditional fully connected neural networks. RNN is known to learn problems with long range temporal dependencies and fast convergence, it makes the encoder efficient enough to capture the dynamic pattern by mapping the sequential data to a highly non-linear latent space. 

In this paper, RNN is applied to preserve the historical information and capture the transitional patterns of dynamic networks in this paper. Extensive empirical experiments on sequential datasets have clearly demonstrated that RNNs with the gating units (e.g. GRU-RNN, LSTM-RNN) can handle long-term dependency problems well and converge fast. Therefore, in order to achieve a more efficient temporal learning, we propose two variations of our model based on the GRU and LSTM respectively for comparison. 

Our MDHNE model uses multiple GRU or LSTM units to efficiently preserve dynamic historical information and capture the evolving pattern of different views. In our MDHNE model, for any RNN, there can be $T$ GRU or LSTM units connected as a chain in it, and cell states and hidden representation are passed in a chain from step l to step $T$ of RNN encoder. The chain propagation mechanism provided an effective way to preserve historical information, and we can use it to encode the network evolution process. 

The input to the encoder RNN is multivariate time series of length $T$, and we feed the input to the GRU or LSTM unit at each time step sequentially and each unit calculates and updates the hidden state over time. Given a series of network views $\{ vie{w_1},vie{w_2},...,vie{w_{|P|}}\} $, we transform the original HIN into a series of homogeneous networks $\{ {G_{{\Phi _1}}},{G_{{\Phi _2}}},...,{G_{{\Phi _{|P|}}}}\} $ corresponding to different network views. Moreover, we setup $|P|$  RNN encoders to perform temporal embedding corresponding to 
$\{ vie{w_1},vie{w_2},...,vie{w_{|P|}}\} $ respectively. 

For a dynamic homogeneous network from $vie{w_k}$ , denoted as $G_{{\Phi _k}}^{}$ , we first slice it into $T$ snapshots $\{ G_{{\Phi _k}}^1,G_{{\Phi _k}}^2,...,G_{{\Phi _k}}^T\} $ and $T$ represent the window size, and the corresponding $k$-th RNN is selected to encode temporal inputs. Specifically, for any node
${v_i}$ , the corresponding temporal input vectors of the corresponding $k$-th RNN from time step 1 to step $T$ can be denoted as $M_{{\Phi _k}}^{}(i,:){\rm{ = \{ }}M_{{\Phi _k}}^1(i,:),M_{{\Phi _k}}^2(i,:),...,M_{{\Phi _k}}^T(i,:){\rm{\} }}$  respectively, which represent different linkage state between ${v_i}$  and other nodes guided by meta-path ${\Phi _k}$  through a series of timestamps. And $M_{{\Phi _k}}^t(i,j)$  represents the proximity between ${v_i}$  and ${v_j}$ through the meta-path ${\Phi _k}$ at time step $t$ calculated by PathSIM\cite{pathsim2011sun} (a SOTA meta-path-based measure). 

Usually, GRU or LSTM units are seen as black boxes. The GRU unit structure can be illustrated below and the relationships between internal parameters are defined as follows:
\begin{equation}\label{MD}
z_{{\Phi _k}}^t = \sigma ({W_z}M_{{\Phi _k}}^t(i,:) + {U_z}h_{{\Phi _k}}^{t - 1}(i)),
\end{equation}
\begin{equation}\label{MD}
r_{{\Phi _k}}^t = \sigma ({W_r}M_{{\Phi _k}}^t(i,:) + {U_r}h_{{\Phi _k}}^{t - 1}(i)),
\end{equation}
\begin{equation}\label{MD}
\tilde h_{{\Phi _k}}^t = \tanh (WM_{{\Phi _k}}^t(i,:) + U(r_{{\Phi _k}}^t \odot h_{{\Phi _k}}^{t - 1}(i)),
\end{equation}
\begin{equation}\label{MD}
h_{{\Phi _k}}^t(i) = (1 - z_{{\Phi _k}}^t) \odot h_{{\Phi _k}}^{t - 1}(i) + z_{{\Phi _k}}^t \odot \tilde h_{{\Phi _k}}^t,
\end{equation}
where $f$ is a non-linear activation function, $M_{{\Phi _k}}^t(i,:)$ denotes as the input vector of current time slice, and $h_{{\Phi _k}}^{t - 1}(i)$ represents the hidden state obtained from the previous time slice. And we merge two inputs and learn the current hidden state $h_{{\Phi _k}}^t(i)$  by current GRU unit. 

Moreover, the hidden state representation of a single LSTM unit can be defined as:

\begin{equation}\label{MD}
f_{{\Phi _k}}^t = \sigma ({W_f} \cdot [h_{{\Phi _k}}^{t - 1}(i),M_{{\Phi _k}}^t(i,:)] + {b_f}),
\end{equation}
\begin{equation}\label{MD}
i_{{\Phi _k}}^t = \sigma ({W_i} \cdot [h_{{\Phi _k}}^{t - 1}(i),M_{{\Phi _k}}^t(i,:)] + {b_i}),
\end{equation}
\begin{equation}\label{MD}
o_{{\Phi _k}}^t = \sigma ({W_o} \cdot [h_{{\Phi _k}}^{t - 1}(i),M_{{\Phi _k}}^t(i,:)] + {b_o}),
\end{equation}
\begin{equation}\label{MD}
c_{{\Phi _k}}^t = f_{{\Phi _k}}^t \odot c_{{\Phi _k}}^{t - 1} + i_{{\Phi _k}}^t \odot ({W_c} \cdot [h_{{\Phi _k}}^{t - 1}(i),M_{{\Phi _k}}^t(i,:)] + {b_c}),
\end{equation}
\begin{equation}\label{MD}
h_{{\Phi _k}}^t(i) = o_{{\Phi _k}}^t \odot \tanh (c_{{\Phi _k}}^t),
\end{equation}
where $f_{{\Phi _k}}^t$  is the value to trigger the forget gate, $o_{{\Phi _k}}^t$ is the value to trigger the output gate, 
$i_{{\Phi _k}}^t$ represents the value to trigger the update gate of the LSTM, $c_{{\Phi _k}}^t$ represents the new estimated candidate state, and b represents the biases.${W_i},{W_f},{W_o}$ are the weighted matrics and ${b_i},{b_f},{b_o}$ are the biases of LSTM to be learned during training, parameterizing the transformations of the input, forget and output gates respectively. $\sigma $ is the sigmoid function and $ \odot $  stands for element-wise multiplication. 

Our proposed MDHNE model passes the sequential input through RNN. After reading the last input of the sequence, the output of RNN encoder will be the compressed latent representation. After the calculation process of decoder, we can obtain the output $(\hat M_{{\Phi _1}}^T,\hat M_{{\Phi _2}}^T,...,\hat M_{{\Phi _k}}^T)$ as the new structure inference. The goal of our model is minimizing the prediction error so that the structure inference $(\hat M_{{\Phi _1}}^T,\hat M_{{\Phi _2}}^T,...,\hat M_{{\Phi _k}}^T)$ can fit the real linkage state $(M_{{\Phi _1}}^T,M_{{\Phi _2}}^T,...,M_{{\Phi _k}}^T)$.

\subsection{Multi-View Fusion Mechanism}\label{subsecSimilaritymeasure}
After a group of node latent representations from different views have been obtained, an efficient fusion mechanism is needed to integrate these latent vectors and further vote for the final node representations. A high-dimension representation can be directly concatenated by all latent representations. Alternatively, all implicit vectors are averaged (i.e., assigning the same weight to all latent vectors). 
Considering that different views make different contributions to the final network embedding, an attention mechanism is introduced in the MDHNE framework so that weights of latent vectors that are encoded in node proximities of different views can be automatically learned. Afterwards, the attention mechanism was adopted to design objective functions in consistency with particular data mining tasks by providing a few label data. 

Specifically, for all latent representations $ \{h_{{\Phi_1}}^T(i)$, $ h_{{\Phi_2}}^T(i)$, $ ... $, $h_{{\Phi_K}}^T(i)\} $ of node 
$v_i^{} \in V$  that are obtained from hidden layers of all corresponding RNN encoders, a two-layer neural network was selected to calculate attention weights. Moreover, the attention weight $att_k^{}$  corresponding to the $k$-th vector can be calculated using a softmax unit as follows:
\begin{equation}\label{MD}
att_k = {h^T} \cdot {\mathop{\rm Tanh}\nolimits} (W \cdot h_{{\Phi _k}}^T(i) + b),
\end{equation}
\begin{equation}\label{MD}
att_k^{} = \frac{{\exp (att_k^{})}}{{\sum\nolimits_{k = 0}^K {\exp (att_k^{})} }},
\end{equation}
where $W$, $b$ and $h$ respectively stand for a weight matrix, a bias vector and a weight vector. Apparently, the greater $att_k^{}$  is, the more important the kth view will be to vote for the final embedding vector. In the proposed embedding framework of this paper, a higher weight means that the corresponding vector contains richer structural information and implicit semantic information in a heterogeneous network. 

After acquiring all attention weights denoted by $att_i^k$ , the learned weights are used as coefficients to carry out combined weight solution for $K$ sets of node latent representations. The final embedding vector of node $v_i^{} \in V$  can be expressed as: 
\begin{equation}\label{MD}
z_i^{} = \sum\limits_{k = 1}^K {at{t_i} \times } h_{{\Phi _k}}^T(i).
\end{equation}

\subsection{Loss Function}\label{subsecSimilaritymeasure}
We propose a weighted combination ${L_{all}}$ of two objectives 
${L_{s{\rm{tructure}}}}$  and ${L_{{\rm{attention}}}}$, which can be denoted as ${L_{all}} = {L_{s{\rm{tructure}}}} + \beta {L_{{\rm{attention}}}}$  , where $\beta $ is the hyper-parameter appropriately chosen as relative weights of the objective functions. We define the structure loss denoted as ${L_{s{\rm{tructure}}}}$ to describe the deviation between inference and actual linkage state, moreover, ${L_{{\rm{attention}}}}$  is the objective function for weight learning of attention mechanism. By minimizing the overall loss function of the collaboration framework, MDHNE can capture the highly nonlinear interaction proximities and transition patterns encoded in different views simultaneously, and meanwhile integrate them to vote for the robust node representations via attention mechanism.   

Next, we will introduce the details of two loss functions respectively: 

\emph{1.The loss function} ${L_{s{\rm{tructure}}}}$: 

As a well known assumption in dynamic network realms demonstrated, each node has a unique transitional pattern through time slices. By mapping the relevant information to latent space, the encoder has the exponential capability to capture non-linear variance. Furthermore, by minimizing the structure loss, we can use the transition information to update embedding and infer structure of the new network. In this way, the transformation patterns over time can be captured, and the structural evolution information of each view can be preserved in the specific node representations. 

Given the inputs of previous ($T - 1$)  time steps, our model passes the sequential data through the $k$-th RNN which corresponds to $vie{w_k}$ $(1 \le k \le K)$. After a series of non-linear mapping functions, the output of RNN encoder $\hat M_{{\Phi _k}}^T$  will be the updated structure inference at the time step $T$. The goal of loss function ${L_{s{\rm{tructure}}}}$  is minimizing the prediction error so that the structure inference $\hat M_{{\Phi _k}}^T$ can fit the practical linkage state $M_{{\Phi _k}}^T$ well. And we choose cross entropy as the loss function, then the loss function under 
$vie{w_k}$  can be expressed as follows: 

\begin{equation}\label{MD}
\begin{aligned}
L_{s{\rm{tructure}}}^k &=  - \sum\limits_{i = 1}^N {M_{{\Phi _k}}^T(i,:)\log } \hat M_{{\Phi _k}}^T(i,:) \\ &=  - \sum\limits_{i = 1}^N {\sum\limits_{j = 1}^N {M_{{\Phi _k}}^T(i,j)} \log } \hat M_{{\Phi _k}}^T(i,j).
\end{aligned}
\end{equation}

By minimizing the predicted loss, we can learn the node representations which contain historical transformation patterns well. However, due to the sparsity of dynamic network, such a supervised learning process cannot be directly applied to our problem. In other words, the number of zero elements in node proximity matrix is far greater than that of nonzero elements. So a traditional loss function cannot be directly applied in network embedding. To solve such a problem, a punishment mechanism is exerted on nonzero elements and then our model may pay more attention to these nonzero elements and define priority while reconstructing them. 

Then the modified loss function can be expressed as follows: 
\begin{equation}\label{MD}
\begin{aligned}
L_{s{\rm{tructure}}}^k &=  - \sum\limits_{i = 1}^N {Z(i,:) \odot M_{{\Phi _k}}^T(i,:)\log } \hat M_{{\Phi _k}}^T(i,:) \\&=  - \sum\limits_{i = 1}^N {\sum\limits_{j = 1}^N {Z(i,j)} M_{{\Phi _k}}^T(i,:)\log } \hat M_{{\Phi _k}}^T(i,:),
\end{aligned}
\end{equation}
where, $ \odot $  refers to Hadamard product. Moreover, if $M_{{\Phi _k}}^T(i,j) = 0$,$Z(i,j) = 1$ , otherwise, $Z(i,j) = \alpha  > 1$ , and $\alpha $ represents sparsity penalty factor. A higher value of $\alpha $  signifies that a higher level of punishment is exerted on non-zero elements. 

Finally, the first to the $K$-th RNN encoders are used to implement fitting for $\{ M_{{\Phi _1}}^T,M_{{\Phi _2}}^T,...,M_{{\Phi _K}}^T\} $  synchronously, and the total loss function for these RNN encoders can be summarized as: 

\begin{equation}\label{MD}
{L_{s{\rm{tructure}}}} = \sum\limits_{k = 1}^K {L_{s{\rm{tructure}}}^k}.
\end{equation}

\emph{2.The loss function} ${L_{{\rm{attention}}}}$: 

In our opinions, relevant parameters including the fusion function are optimized according to specific data mining tasks. 
For the node classification task, we can minimize the cross entropy over all labeled node between the ground truth and the prediction:

\begin{equation}\label{MD}
{L_{attention}} =  - \sum\limits_{i \in {o_L}} {{o_i}\log (\omega {z_i})},
\end{equation}
where $\omega $  represents parameters of the classifier, ${o_L}$  is the set of node indices that have labels, and ${o_i}$  stands for labels of ${z_i}$ . With the guide of labeled data, we can optimize the proposed model via back propagation and learn the embeddings of nodes. 

In addition, for the recommendation task, the labeled instances are a collection of practical links between nodes which belong to different two types, such as viewing relationships between user nodes and movie nodes in MovieLens, and purchasing relationships between user nodes and item nodes in Amazon. For any node pair $v_i^A$ and $v_i^B$  which respectively belongs to type A and B, after a series of temporal encoding from different views, we can obtain the aggregated latent representations of them, denoted as$(z_i^A,z_j^B)$. Then the probability of the interaction between $(z_i^A,z_j^B)$ can be calculated as follows:

\begin{equation}\label{MD}
{\hat y_{ij}} = sigmoid(z_i^A,z_j^B) = \frac{1}{{1 + {e^{ - z_i^A*z_j^B}}}},
\end{equation}
where the sigmoid(.) is the sigmoid layer, and ${\hat y_{ij}}$  is the probability in the range of [0, 1]. 

Then, the loss function of our model is a point-wise loss function in equation below:

\begin{equation}\label{MD}
{L_{attention}} =  - \!\!\!\sum\limits_{i,j \in Y \cup {Y^ - }} \!\!\!{({y_{ij}}\log {{\hat y}_{ij}} + (1 - {y_{ij}})\log (1 - {{\hat y}_{ij}}))}, 
\end{equation}
where ${y_{ij}}$  is the ground truth of the label instance and the Y and the Y- are the positive and negative instances set, respectively.

\subsection{Model Training}\label{subsecSimilaritymeasure}

By combining the stochastic gradient descent (SGD) and Adaptive Moment Estimation (Adam), relevant parameters including structure loss function and interaction loss function can be continuously optimized. We first make a forward propagation to calculate the loss and then back propagate based on the minimizing the loss function, then relevant model parameters and weights that correspond to different views can be automatically and continuously updated in each iteration. Here, only a few label data corresponding to specific mining task is needed to train the attention mechanism and fine-tune the RNN encoder. With the above learned weights as coefficients, different view-specific node representations can be weighted combined to vote for the robust and accurate representations.

\section{Experimental evaluation}\label{sec:Experimental evaluation}

In this section, we empirically evaluate the effectiveness of the MDHNE method on dynamic HIN. Three real-world dynamic HIN datasets MovieLens, Amazon and DBLP are introduced in the first place. Then we briefly introduce baseline methods for comparison. Subsequently, effectiveness of the proposed MDHNE model is analyzed according to network data mining tasks. Next, we first introduce the related experiment setup before presenting analysis of the experimental results.

\subsection{Experiment Setting}\label{subsecSimilaritymeasure}
\subsubsection{Datasets}
\label{sec:encoder}

To verify the effectiveness of the dynamic heterogeneous embedding framework proposed in this paper, we selected three real-world dynamic network datasets, DBLP, Movielens and Amazon. The concrete description of these heterogeneous information networks is shown in the following table below. (The detailed statistics of these datasets are listed in Table 1.)

\begin{table}
	\centering
	\renewcommand{\thetable}{\Roman{table}}
	\caption{Architectures of models for ablation analysis.}
	\begin{tabular}{|c|c|c|c|c|} 
		\hline 
		Dataset & Relation &  A &  B &  A-B \\
		\hline  
		\multirow{3}*{DBLP}     
		&Paper-Author &25473 &28241 &70652\\
		
		&Paper-Venue &25473 &18 &25437\\
		
		&Paper-Term &25473 &11230 &152105\\
		\hline
		\multirow{5}*{ MovieLens} 
		
		&User-Movie &2106 &10197 &37555 \\
		
		&Movie-Actor &10197 &95241 &231737 \\
		
		&Movie-Director &10197 &4053 &10197\\
		
		&Movie-Tag &10197 &13222 &51795\\
		
		&Movie-Genres &10197 &20 &20809\\
		\hline
		
		\multirow{4}*{ Amazon }
		
		&User-Item &6084	&2753	&195791 \\
		
		&Item-Brand 	&2753	&334	&2753\\
		
		&Item-Category &2753	&22	&5508\\
		
		&Item-View &2753	&3857	&5694\\
		
		\hline
	\end{tabular}
	\label{tab:model-arch}
\end{table}

DBLP: DBLP is an academic network dataset in the field of computer science. DBLP-4area, a data sub-set extracted from DBLP, contains relevant literature information of four research areas: databases, information retrieval, data mining and machine learning. Moreover, such a dataset involves four node types (paper, author, venue and keyword) and their link information such as author-paper (A-P), paper-venue (P-V) and paper-keyword (P-T) relations. The corresponding network schema has been presented in Fig. 1(a). The dataset contains 16405 papers published from year 2000 to 2009, each paper is linked to a publication date, with a granularity of year. In this paper, papers in 10 years (2000-2009) are selected as ten snapshots in the experiment. Each snapshot contains the network structure of one years.

MovieLens: This dataset comprises film viewing records of massive users and other details related to movies. A movie sub-set consisting of five cinematic genres such as action, adventure, science and education and crime was extracted. And each movie falls into at least one of these genres. Then, a heterogeneous network was constructed. Nodes of four types (movie, actor, director and genre) together with link information among them are included in it. The corresponding network schema has been given in Fig. 1(b). and the number of snapshots is set to 6 from 2003 to 2008, we also fix the time difference between network snapshots to one year.

Amazon: This dataset records user rating on businesses and contains social relations and attribute data of businesses. In our experiment, we select the items of Electronics categories for evaluation. Moreover, such a dataset involves four node types (user, item, brand, view and category) and their link information such as user-item (U-I), item-category (I-C) and item-brand (I-B) relations. In this paper, the Amazon dataset contains about 6k users and 2.7k items, with user-item rating timestamps ranging from 2000 to 2013, data ranging from 2000 to 2013 is selected as fourteen snapshots in the experiment, and we fix the time difference between network snapshots to one year.

These three real-world datasets also have different linkage sparsity degrees: the DBLP is sparse, while the MovieLens and Amazon is relatively denser.

\subsubsection{Baselines for Comparison}
\label{sec:encoder}

Various state-of-the-art methods are applied as the baselines to compare with our MDHNE framework in this paper. We make a full comparison with these baselines to show the capability of our proposed method in node classification task and recommendation task respectively.

DeepWalk\cite{perozzi2014deepwalk}: a static homogeneous network embedding approach. This method first applies random walk to generate sequences of nodes from the network, and then uses it as input to the skip-gram model to learn representations.

Node2vec\cite{grover2016node2vec}: a static homogeneous network embedding approach. Node2vec defines two parameters of p and q, so that random walk strikes a balance between BFS and DFS to preserve local and global information of nodes. Therefore, this model is greatly adaptive, and in this paper the experimental parameter is defined as p=q=1.0.

Metapath2vec\cite{dong2017metapath2vec}: a static heterogeneous network embedding approach. Random walk is generated under the meta-path constraints to construct a neighbor node set of nodes and then acquire node vector representation based on the heterogeneous skip-gram model. Considering that HIN may have diverse meta-paths, we select the most efficient meta-path for experiment here to guide random walk.

HIN2Vec\cite{fu2017hin2vec}: a static heterogeneous network embedding approach. The core of the HIN2Vec framework is a neural network model, which is designed to capture the rich semantics embedded in HINs by exploiting different types of relationships among nodes. We can learn the latent representations of nodes and meta-paths in an HIN by conducting multiple prediction training tasks jointly.

dynGEM\cite{goyal2018dyngem}: a dynamic homogeneous network embedding approach, which utilizes the deep auto-encoder to incrementally generate dynamic embedding at the time step t by using only the snapshot at the t-1 time slice.

dyngraph2vecRNN\cite{goyal2020dyngraph2vec}: a dynamic homogeneous network embedding approach, which uses sparsely connected Long Short Term Memory (LSTM) networks to learn the embedding.

MDHNEavg: A variant of our proposed MDHNE model. In this model, we cancel the attention mechanism and fuse the learned embeddings from different views with the same weight coefficient.

MDHNElstm and MDHNEgru: Two version of our proposed MDHNE model, which apply LSTM or GRU based RNN respectively to incorporate the transformation patterns of dynamic HINs to update the embedding results incrementally.

For static embedding models above, Node2vec, Metapath2vec and Metagraph2vec approaches can only handle static networks. To have a fair comparison with our proposed MDHNE framework, we rerun these static embedding independently at each snapshot and the average performance over all time steps are reported. Another two dynamic embedding methods dyngraph2vecRNN and DynGem are designed for dynamic homogeneous network, but heterogeneity of nodes and links are neglected. 

Moreover, for the recommendation task, following recommendation approaches are also considered as baselines and the brief introductions of them are listed as follows: 

DMF\cite{xue2017deep}: DMF uses the interaction matrix as the input and maps users and items into a common low-dimensional space using a deep neural network.

HERec\cite{shi2018heterogeneous}: It utilizes a meta-path based random walk strategy to generate meaningful node sequences for network embedding. The learned node embeddings are first transformed by a set of fusion functions, and subsequently integrated into an extended Matrix Factorization (MF) model for the rating prediction task.

\subsubsection{Parameters Settings}
\label{sec:decoder}

The experimental environment is described as follows: All codes of proposed methods are performed on a Linux server with ubuntu 16.04 operating system and programming platform of Python3.6 + tensorflow-gpu 1.2.0. The server is equipped with 32G RAM, Intel Xeon E5-2690 central processing units (CPU), and double Nvidia GTX-2080Ti (GPU). 

The core of the proposed MDHNE model is the RNN-encoder which is sequential neural network with multiple GRU or LSTM units. In the training stage, we randomly initialize the model parameters with a Gaussian distribution. Here, label data are selected to fine-tune the RNN-encoder and automatically learn weights of proximities from different views, and batch size of stochastic gradient descent is set as 500, and original learning rate is set as 0.005. We optimize all models with Adam optimizer, and adopt the default Xavier initializer to initialize the model parameters. Among all aforementioned baseline approaches, HIN based methods need to specify the used meta-paths. We only select short meta-paths of at most four steps, since long meta-paths are likely to introduce noisy semantics. For the sake of fair comparison, the dimension of node representations learned by all embedding models in the experiment is uniformly set as 128. For different datasets, the parameters of the baseline are different, all parameters are fine-tuned to the optimal. And for all baseline method, we perform a grid search on the values of hyper parameters, and we choose a specific combination of them for each task on each dataset. 

\subsection{Node Classification Task}\label{subsecSimilaritymeasure}

We start by evaluating the quantitative results through the node classification task. Node representations, which are learned on DBLP dataset from our proposed model and baseline embedding approaches, are treated as features in the node classification task. Relevant classification results are assessed to estimate updated embedding results are good or not. GBDT is chosen as the classifier algorithm using the sklearn package in python. In the course of experiment, training set is set by updated embedding results of current time step, and percent occupied by the training set is randomly sampled to 90 from 10. Node representations in the training set are used to train the classifier GBDT which is then used on the testing set. At each time step, we say the label of each author is the corresponding conferences that his/her papers were mainly published in. The experiment is repeated for 10 times and the average experimental results are reported. Micro-F1 and Macro-F1 are selected as evaluation metrics of this task. We present the results of different approaches on the node classification task in Table 2. The highest score of each group was marked by boldfaced characters. 

\begin{table}
	\normalsize
	\setlength{\abovecaptionskip}{0cm}
	\setlength{\belowcaptionskip}{0.2cm}
	\centering
	\caption{Performance on node classification of DBLP.}
	\label{Performance of node similarity search}
	\setlength{\tabcolsep}{2mm}
	\scalebox{0.7}
	{
		\begin{tabular}{|c|c|c|c|c|c|c|} 
			\hline
			
			Method & metrics & $10\%$ & $30\%$ & $50\%$ & $70\%$ & $90\%$\\
			\hline
			\multirow{2}{*}{Node2vec}
			& Micro-F1 & 0.5233 & 0.5415 & 0.5482 & 0.5582 & 0.5619\\
			
			& Macro-F1  &0.5115 &0.5307  &0.5387 & 0.5562  & 0.5538 \\
			\hline
			\multirow{2}{*}{Metapath2vec} 
			& Micro-F1 & 0.5596	&0.5868	&0.6048	&0.6199		&0.6203\\
			
			& Macro-F1	 &0.5462	&0.5783	&0.5984	&0.6070	&0.6148 \\
			\hline
			\multirow{2}{*}{HIN2Vec} 
			& Micro-F1 & 0.6798	 &0.6888 &	0.6999 &	0.7171  &	0.7192\\
			&Macro-F1	 &0.6703	&0.6829	&0.6910	&0.7137	&0.7151 \\
			\hline	
			\multirow{2}{*}{DynGEM}
			& Micro-F1 & 0.6030	&0.6344 &	0.6453	&0.6560	&0.6615\\
			& Macro-F1	 &0.5955	&0.6282	&0.6399	&0.6579	&0.6551 \\
			\hline	
			\multirow{2}{*}{dyngraph2vec}
			& Micro-F1 & 0.6619	& 0.6714	& 0.6795	& 0.6908		& 0.6900\\
			& Macro-F1	 &0.6664 &	0.6770	&0.6858		&0.6976	&0.6980 \\
			\hline
			\multirow{2}{*}{MDHNEavg}
			& Micro-F1 & 0.7083	&0.7116	&0.7138	&0.7302		&0.7318\\
			& Macro-F1	 &0.6927	&0.6943	&0.7023	&0.7188	&0.7242 \\
			\hline
			
			\multirow{2}{*}{MDHNEgru}
			& Micro-F1 & 0.7150	& 0.7511	& 0.7610	& 0.7685		& 0.7696\\
			& Macro-F1	 &0.7033 &	0.7439	&0.7536		&0.7514	&0.7515 \\
			\hline
			\multirow{2}{*}{MDHNElstm}
			& Micro-F1 & 0.7090	&0.7498	&0.7647	&0.7650		&0.7611\\
			& Macro-F1	 &0.7089	&0.7431	&0.7625	&0.7510	&0.7540 \\
			\hline
		\end{tabular}
	}
	
\end{table}   

As the classification results show, our method shows superior performance than other baselines on DBLP with all varying sizes of training data, as measured by both micro-F1 and Macro-F1. The stable performances of our methods against different training sizes indicate the robustness of our learned node representations when served as features for node classification task. The traditional static homogeneous embedding methods (e.g., Node2vec) which neglect both the heterogeneity of network and the temporal interaction information, give relatively poor classification results. Static HIN embedding models such as Metapath2vec and HIN2Vec, which can leverage complex semantics information, perform better than node2vec. Moreover, though designed for homogeneous network, DynGEM and dyngraph2vec which combine the temporal transformation information perform well. From these experiments, we can conclude that multi-view proximity and incorporating dynamic changes are both of paramount importance in network embedding. Besides, MDHNE performs better than MDHNEavg as measured by macro-F1 and micro-F1, which indicates that attention on different views could actually help to learn better node representations.

\begin{table*}
	\normalsize
	\setlength{\abovecaptionskip}{0cm}
	\setlength{\belowcaptionskip}{0.2cm}
	\centering
	\caption{Comparison on Amazon dataset.}
	\label{Performance of node similarity search}
	\setlength{\tabcolsep}{2mm}
	\scalebox{0.8}
	{
		\begin{tabular}{|c|c|c|c|c|c|c|c|c|} 
			\hline 
			Metrics	&HR@5	&HR@10	&HR@15	&HR@20	&NDCG@5	&NDCG@10 &NDCG@15	&NDCG@20 \\
			\hline  
			Node2vec	&0.0537	&0.1103	&0.1566 &0.2143		&0.0320	&0.0502	&0.0624 	&0.0760	\\
			\hline  
			HIN2vec	&0.0890	&0.1750	&0.2531 &0.3516	&0.0521	&0.0797	&0.1002	&0.1234	\\
			\hline  
			DynGEM	&0.0959	&0.1859	&0.2670 &0.3471	&0.0560	&0.0847	&0.1061 	&0.1296	\\
			\hline  
			dyngraph2vec	&0.1691	&0.2556	&0.3733 &0.5602		&0.1103	&0.1383	&0.1536 	&0.2046	\\
			\hline  
			DMF	&0.1253	&0.2744	&0.4212 &0.5553	&0.0728	&0.1234	&0.1594 	&0.1935	\\
			\hline  
			HERec	&0.1432	&0.2133	&0.3475 &0.5962		&0.0824	&0.1255	&0.1598 	&0.1935	\\
			\hline  
			MDHNEavg	&0.1780	&0.2565	&0.4612 &0.5674		&0.1083	&0.1249	&0.1940 	&0.2256	\\
			\hline  
			MDHNEgru	&0.1886	&0.2883	&0.5132 &0.6366		&0.1135	&0.1389	&0.2139 	&0.2266	\\
			\hline  
			MDHNElstm	&0.1307	&0.2491	&0.5102 &0.6425		&0.0883	&0.1255	&0.1940 	&0.2286	\\
			\hline  
			
		\end{tabular}
	}
	
\end{table*} 

\begin{table*}
	\normalsize
	\setlength{\abovecaptionskip}{0cm}
	\setlength{\belowcaptionskip}{0.2cm}
	\centering
	\caption{Comparison on MovieLens dataset.}
	\label{Performance of node similarity search}
	\setlength{\tabcolsep}{2mm}
	\scalebox{0.8}
	{
		\begin{tabular}{|c|c|c|c|c|c|c|c|c|} 
			\hline 
			Metrics	&HR@5	&HR@10	&HR@15	&HR@20	&NDCG@5	&NDCG@10 &NDCG@15	&NDCG@20 \\
			\hline  
			Node2vec	&0.0276	&0.0905	&0.1474 &0.2044		&0.0343	&0.0512	&0.0623 	&0.0717	\\
			\hline  
			HIN2vec	&0.3586	&0.6308	&0.4836 &0.4922	&0.1198	&0.2065	&0.2235	&0.2339	\\
			\hline  
			DynGEM	&0.1613	&0.2648	&0.4005 &0.4206	&0.0650	&0.1296	&0.1658 	&0.1706	\\
			\hline  
			dyngraph2vec	&0.4179	&0.6256	&0.6984 &0.7010		&0.2828	&0.3491	&0.3680 	&0.3695	\\
			\hline  
			DMF	&0.2646	&0.4505	&0.6080 &0.6390	&0.1649	&0.2242	&0.2661 	&0.2736	\\
			\hline  
			HERec	&0.3886	&0.6608	&0.7236 &0.7252		&0.2698	&0.3565	&0.3735 	&0.3739	\\
			\hline  
			MDHNEavg	&0.4179	&0.6314	&0.7026 &0.7060		&0.2817	&0.3495	&0.3688 	&0.3696	\\
			\hline  
			MDHNEgru	&0.4765	&0.8768	&0.8994 &0.8995		&0.2706	&0.4024	&0.4085 	&0.4085	\\
			\hline  
			MDHNElstm	&0.4715	&0.8827	&0.9065 &0.9112		&0.2693	&0.4050	&0.4124 	&0.4128	\\
			\hline  
			
		\end{tabular}
	}
	
\end{table*}   

\subsection{Recommendation Task}\label{subsecSimilaritymeasure}
A qualified network representation method can not only reconstruct the visible edge in the training, but also predict the edge that should appear but lose in training data. Considering that the research object in this paper is heterogeneous network, there is no direct connection between nodes of the same type, so we choose the recommended task, which is used to test the ability to predict the interaction links between two types of nodes. For recommended task, in Amazon data set, we predict the purchase relationship between user nodes and commodity nodes, and in movielens dataset, we predict the viewing relationship between user nodes and movie nodes. Then we train the embedding vectors on the training network and evaluate the recommendation performance on the testing network. We apply the leave-one-out method for evaluation. For a fair comparion with the baselines, we use the same negative sample set for each (user, item) or (user, movie) pair in the test set of Amazon and MovieLens respectively for all the methods. After that, we adopt the widely-used evaluation protocols: HR@K and NDCG@K metrics to measure the quality of recommendation. We set $k = 5,10,15,20$ , and the average metrics are reported for all users in the test set. The results of recommendation task are reported in Table.3 and Table.4 with HR and NDCG score respectively.

We observe that at least $8.4\% $ and $4.7\% $  improvement in HR@K and NDCG@K values respectively generated by MDHNE when compared with other baselines on amazon. Moreover, in MovieLens dataset, our model improves HR@K and NDCG@K by approximately $17\% $ and $7.8\%$ respectively. The results show that our MDHNE achieves the best among all baselines regardless whether the datasets are sparse or dense. In our proposed method, both heterogeneous semantics information and transition of network are well considered so that the embedding results of MDHNE can be updated to keep freshness. It is worth noting that conventional embedding methods do not perform well, since they can only preserve the structural information but ignore the transitional information of dynamic networks. In addition, our model considers multiple views in terms of users' preferences, and thus can indeed facilitate the updating of embedding and learn better representations of nodes, which is beneficial for predicting interactions between nodes.

\subsection{Parameter Sensitivity Evaluation}\label{subsecSimilaritymeasure}

Sensitivity of the MDHNE framework to parameters is further analyzed, involving (1) Dimension of the learned node vectors, (2) Length of historical snapshots considered for training (3) Impact of different views, (4) Number of views. During the following experiments, a proportion taken by the training set was set at 0.3. Related parameters are modified to analyze embedding performance. 

\begin{figure*}[ht]
	\centering
	\subfigure[HR@$k$ with MDHNEgru.]{
		\begin{minipage}[t]{0.25\linewidth}
			\centering
			\includegraphics[width=1.15\linewidth]{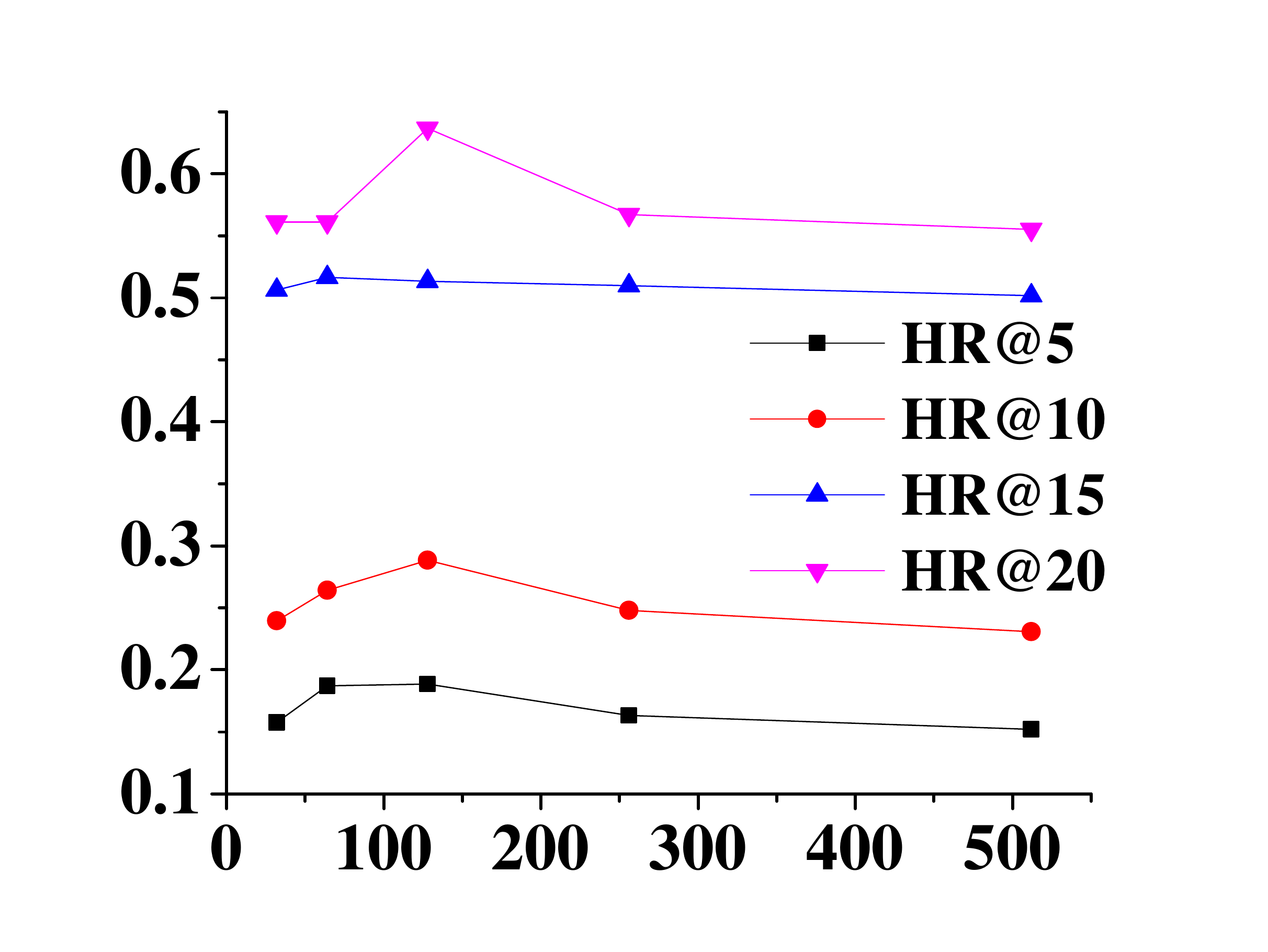}
		\end{minipage}%
	}%
	\subfigure[NDCG@$k$ with MDHNEgru.]{
		\begin{minipage}[t]{0.25\linewidth}
			\centering
			\includegraphics[width=1.15\linewidth]{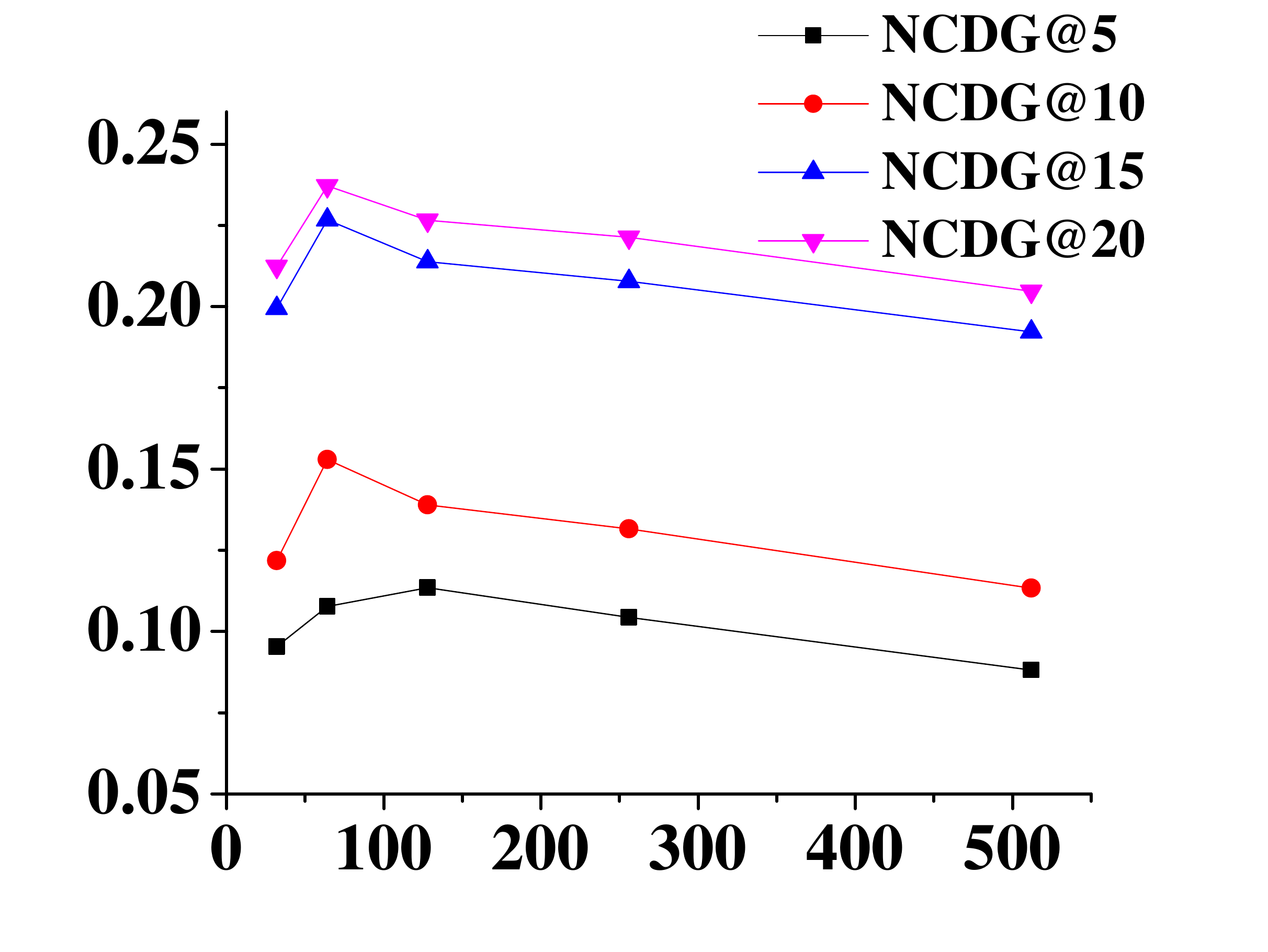}
		\end{minipage}%
	}%
	\subfigure[HR@$k$ with MDHNElstm.]{
		\begin{minipage}[t]{0.25\linewidth}
			\centering
			\includegraphics[width=1.15\linewidth]{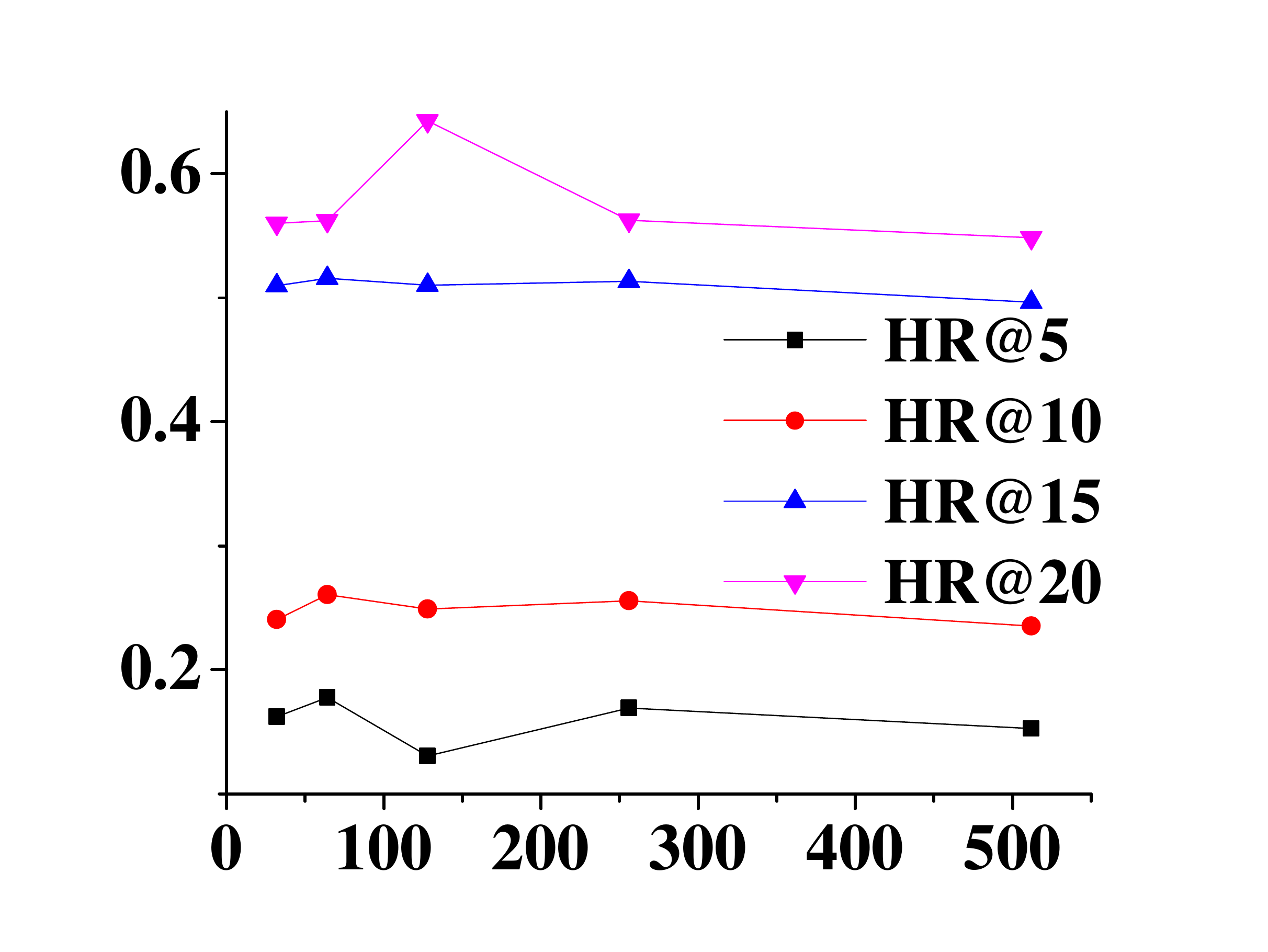}
		\end{minipage}
	}%
	\subfigure[NDCG@$k$with MDHNElstm.]{
		\begin{minipage}[t]{0.25\linewidth}
			\centering
			\includegraphics[width=1.15\linewidth]{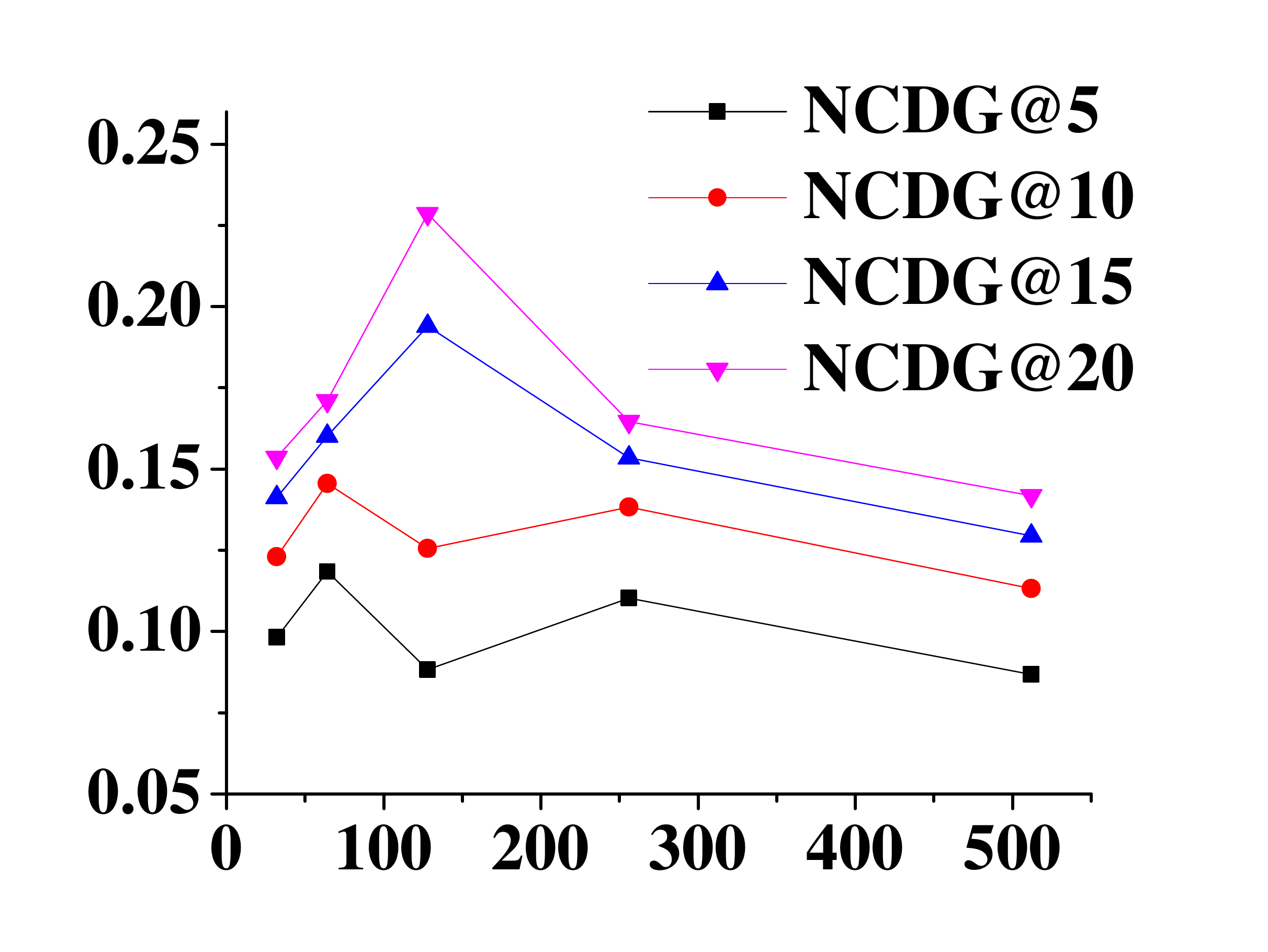}
		\end{minipage}
	}%
	\centering
	\caption{Impacts of dimension on Amazon.}
\end{figure*}

\begin{figure*}[ht]
	\centering
	\subfigure[HR@$k$ with MDHNEgru.]{
		\begin{minipage}[t]{0.25\linewidth}
			\centering
			\includegraphics[width=1.15\linewidth]{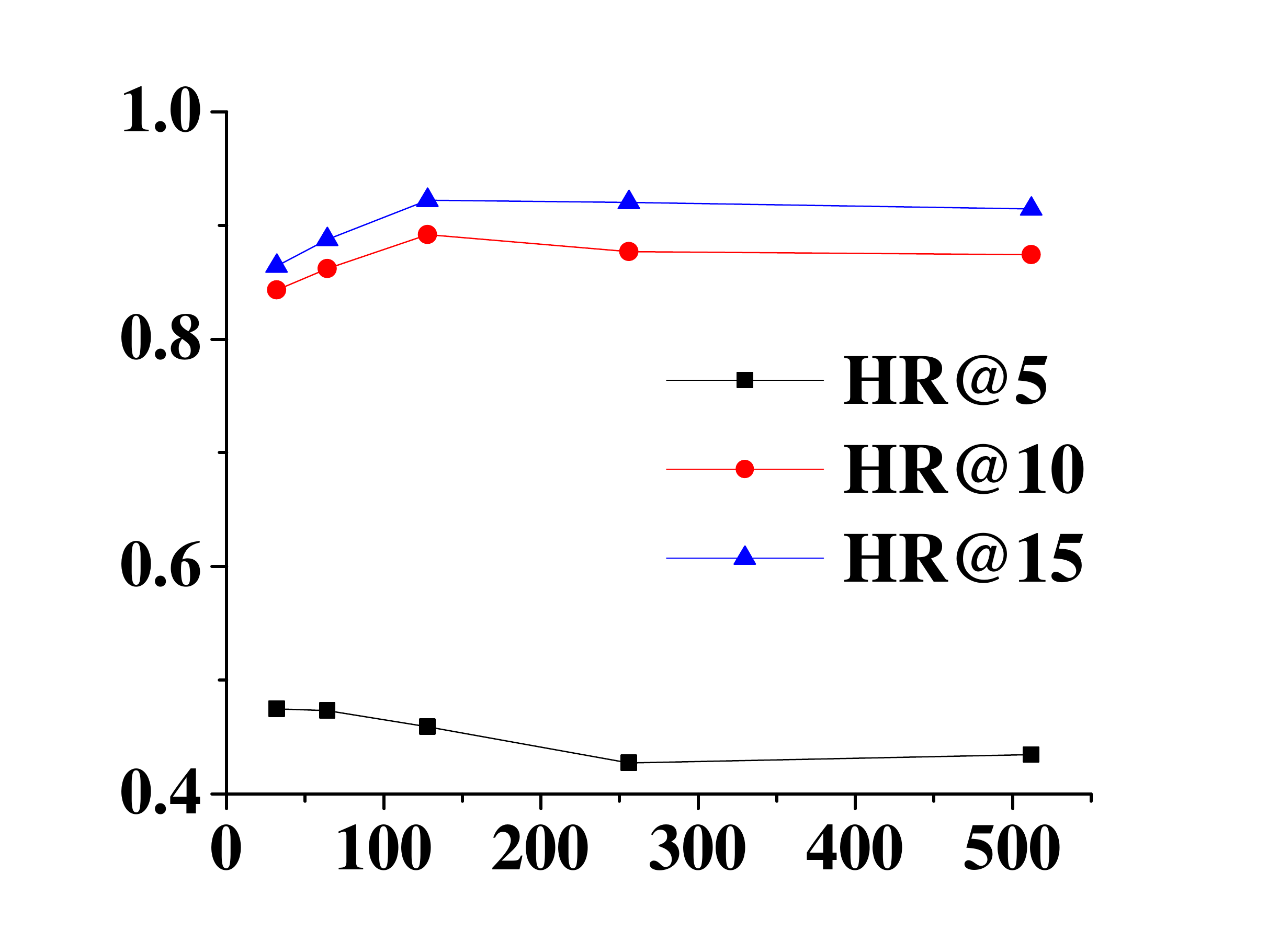}
		\end{minipage}%
	}%
	\subfigure[NDCG@$k$ with MDHNEgru.]{
		\begin{minipage}[t]{0.25\linewidth}
			\centering
			\includegraphics[width=1.15\linewidth]{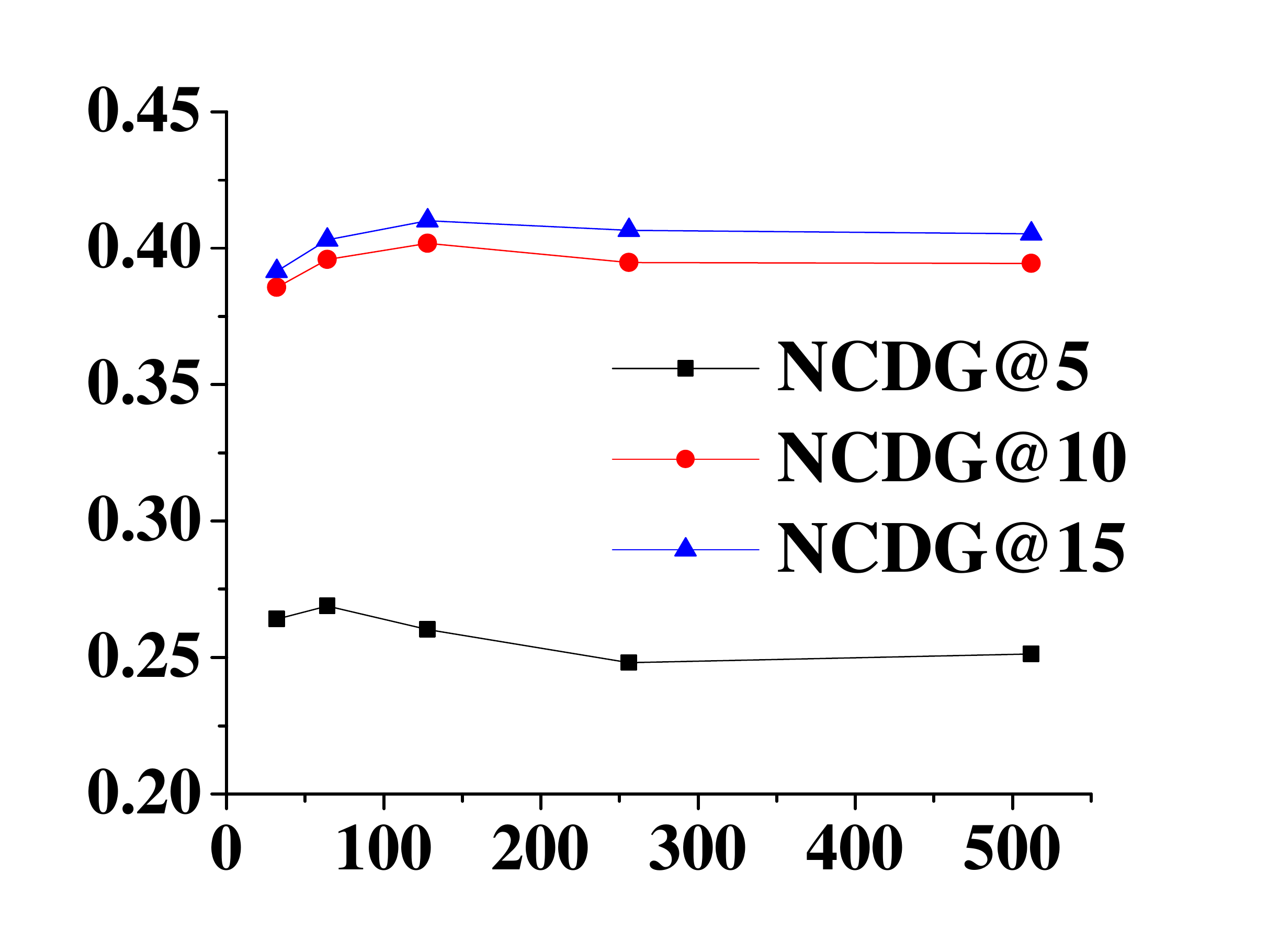}
		\end{minipage}%
	}%
	\subfigure[HR@$k$ with MDHNElstm.]{
		\begin{minipage}[t]{0.25\linewidth}
			\centering
			\includegraphics[width=1.15\linewidth]{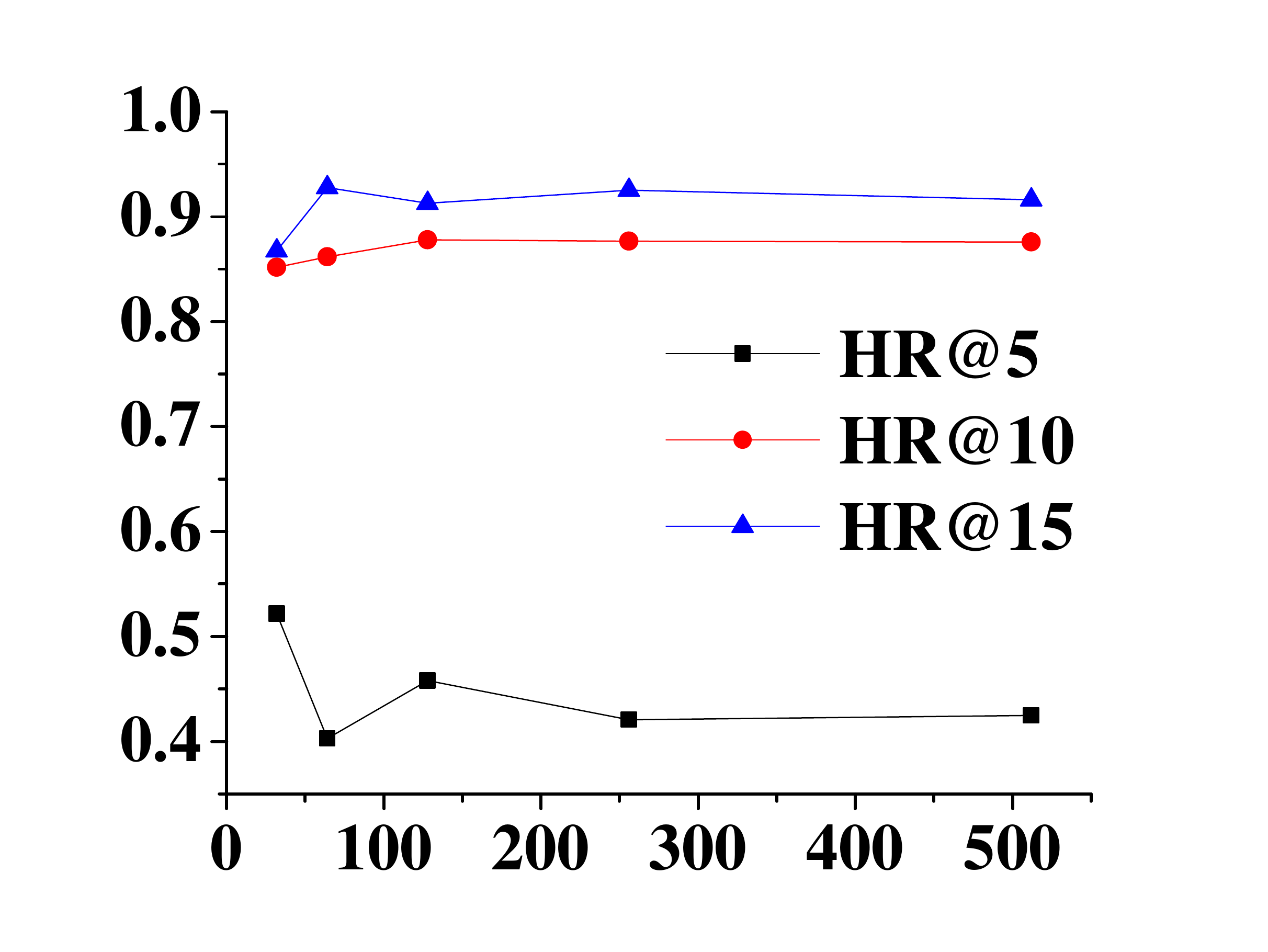}
		\end{minipage}
	}%
	\subfigure[NDCG@$k$with MDHNElstm.]{
		\begin{minipage}[t]{0.25\linewidth}
			\centering
			\includegraphics[width=1.15\linewidth]{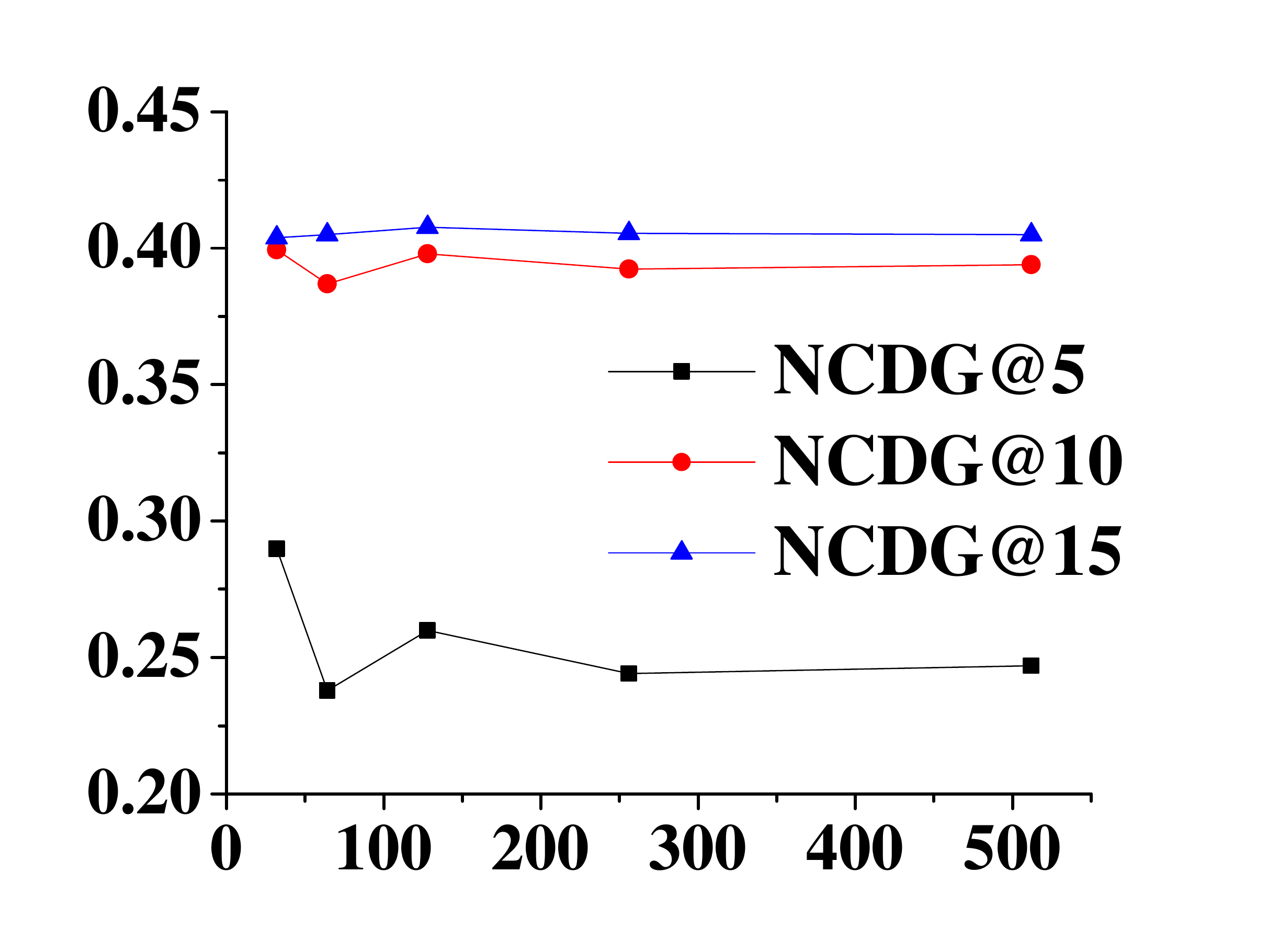}
		\end{minipage}
	}%
	\centering
	\caption{Impacts of dimension on MovieLens.}
\end{figure*}

\begin{figure*}[ht]
	\centering
	\subfigure[HR@$k$.]{
		\begin{minipage}[t]{0.25\linewidth}
			\centering
			\includegraphics[width=1.15\linewidth]{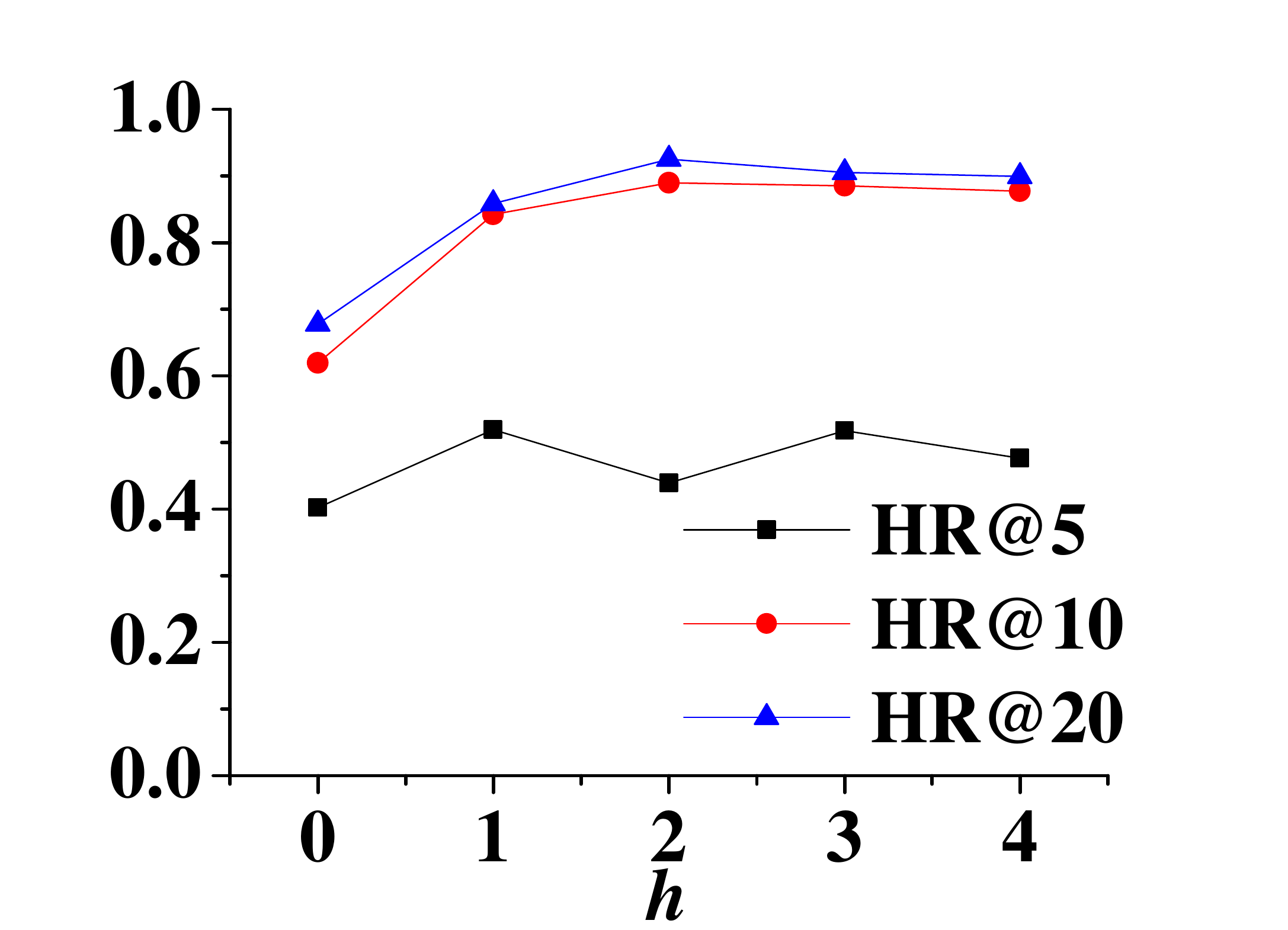}
		\end{minipage}%
	}%
	\subfigure[NDCG@$k$.]{
		\begin{minipage}[t]{0.25\linewidth}
			\centering
			\includegraphics[width=1.15\linewidth]{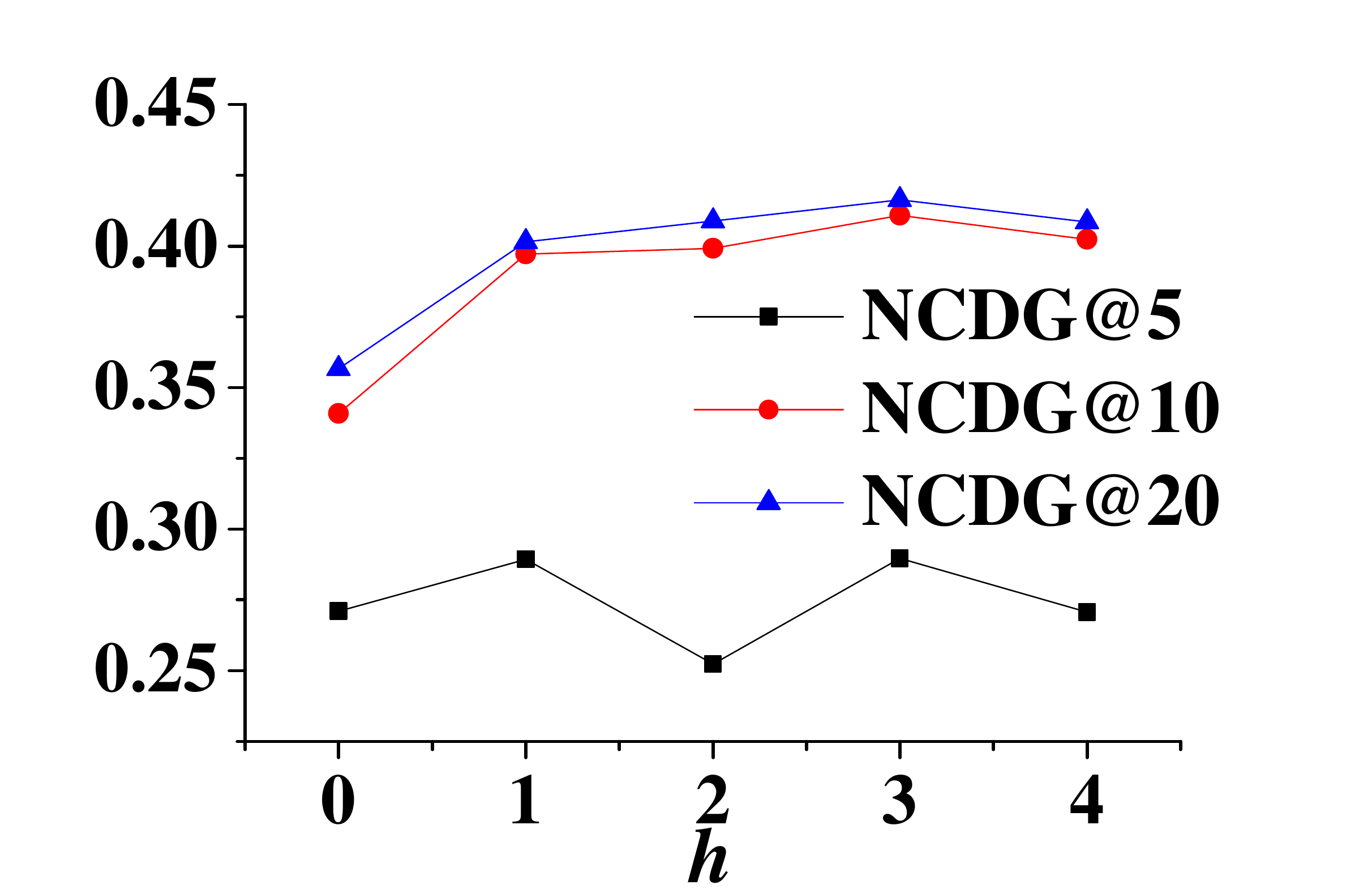}
		\end{minipage}
	}%
	\centering
	\caption{Impacts of dimension on MovieLens.}
\end{figure*} 

\begin{figure*}[ht]
	\centering
	\subfigure[HR@$k$ on Amazon.]{
		\begin{minipage}[t]{0.25\linewidth}
			\centering
			\includegraphics[width=1.15\linewidth]{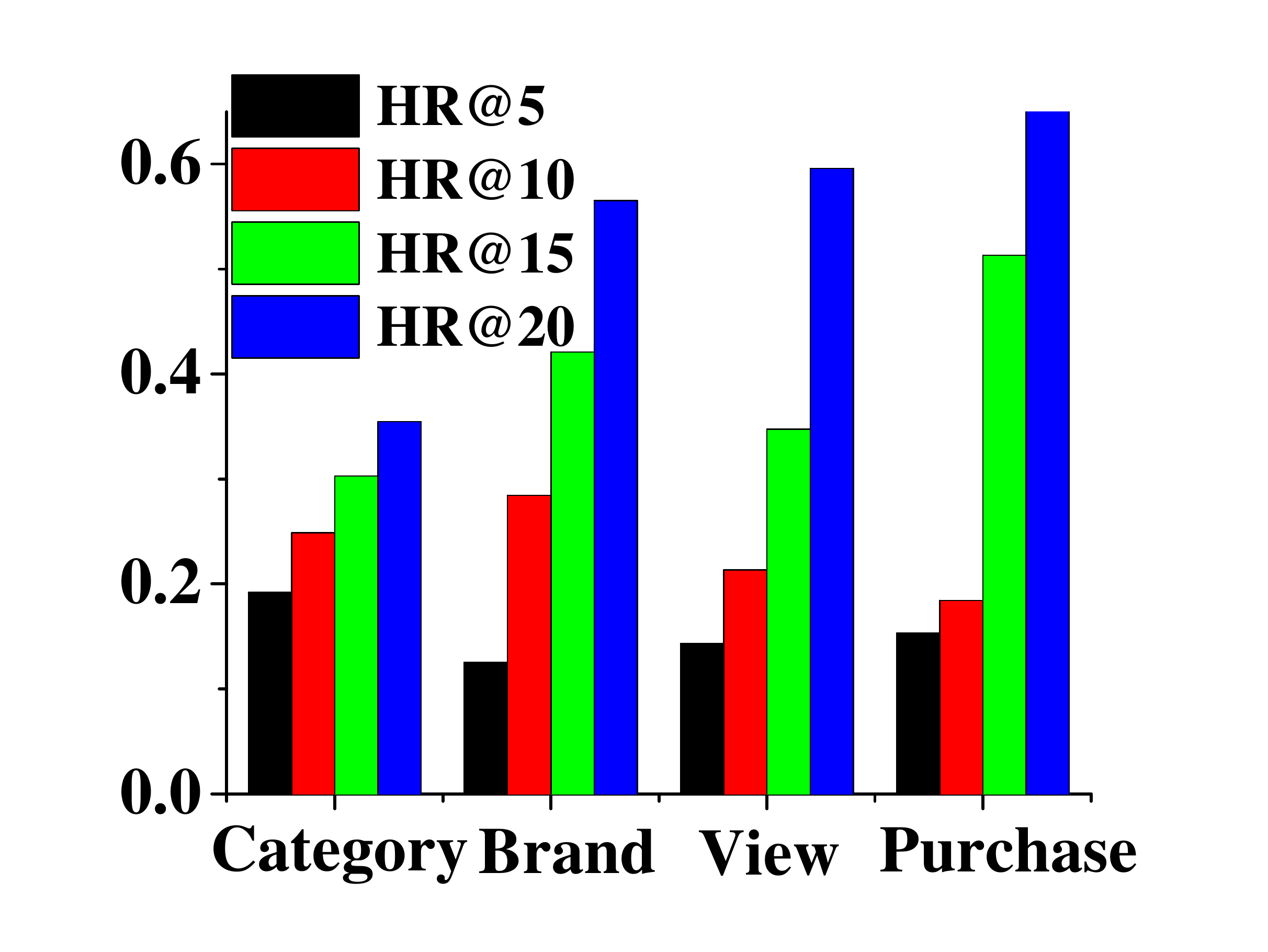}
		\end{minipage}%
	}%
	\subfigure[NDCG@$k$ on Amazon.]{
		\begin{minipage}[t]{0.25\linewidth}
			\centering
			\includegraphics[width=1.15\linewidth]{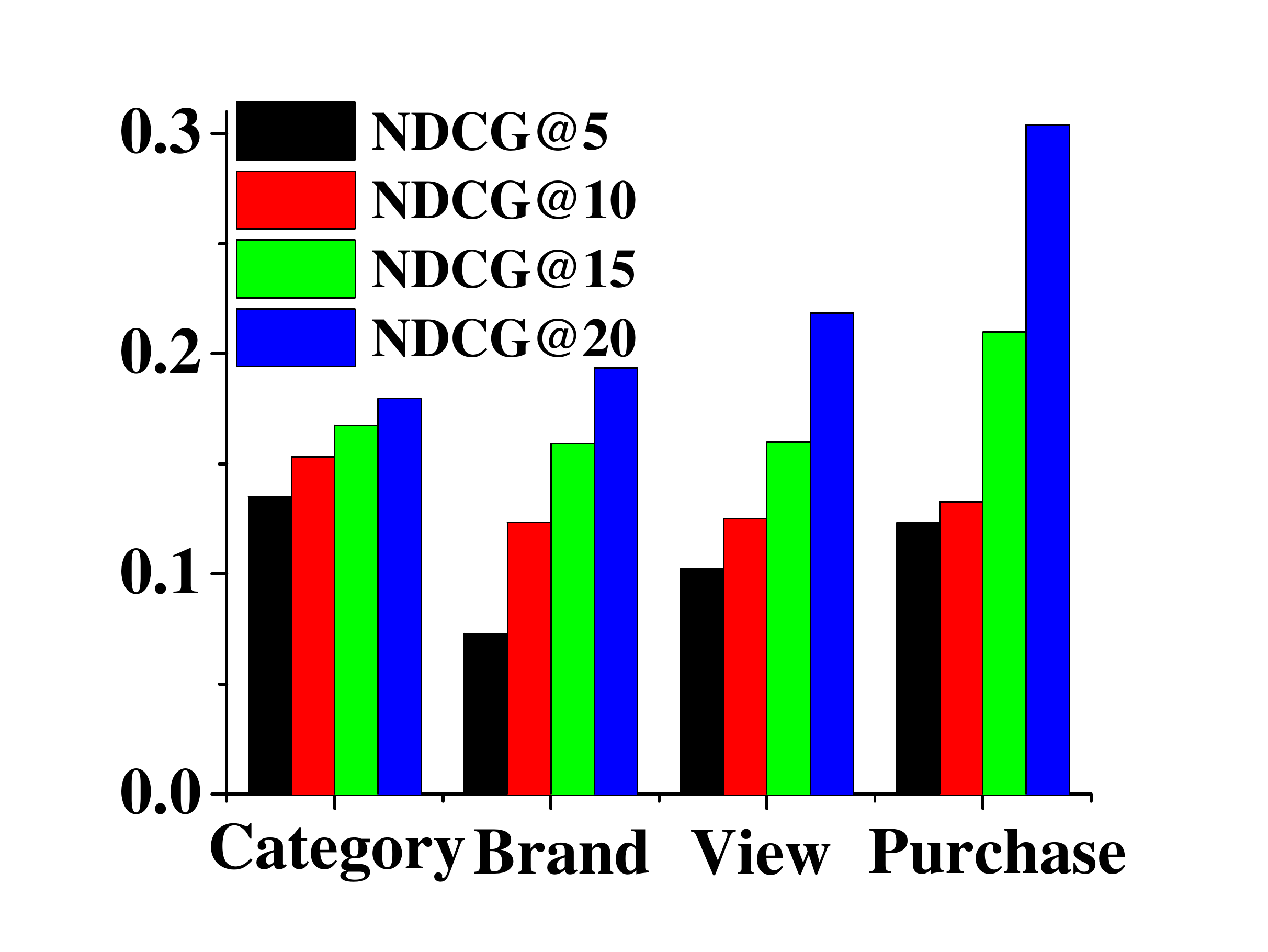}
		\end{minipage}%
	}%
	\subfigure[HR@$k$ on Movielens.]{
		\begin{minipage}[t]{0.25\linewidth}
			\centering
			\includegraphics[width=1.15\linewidth]{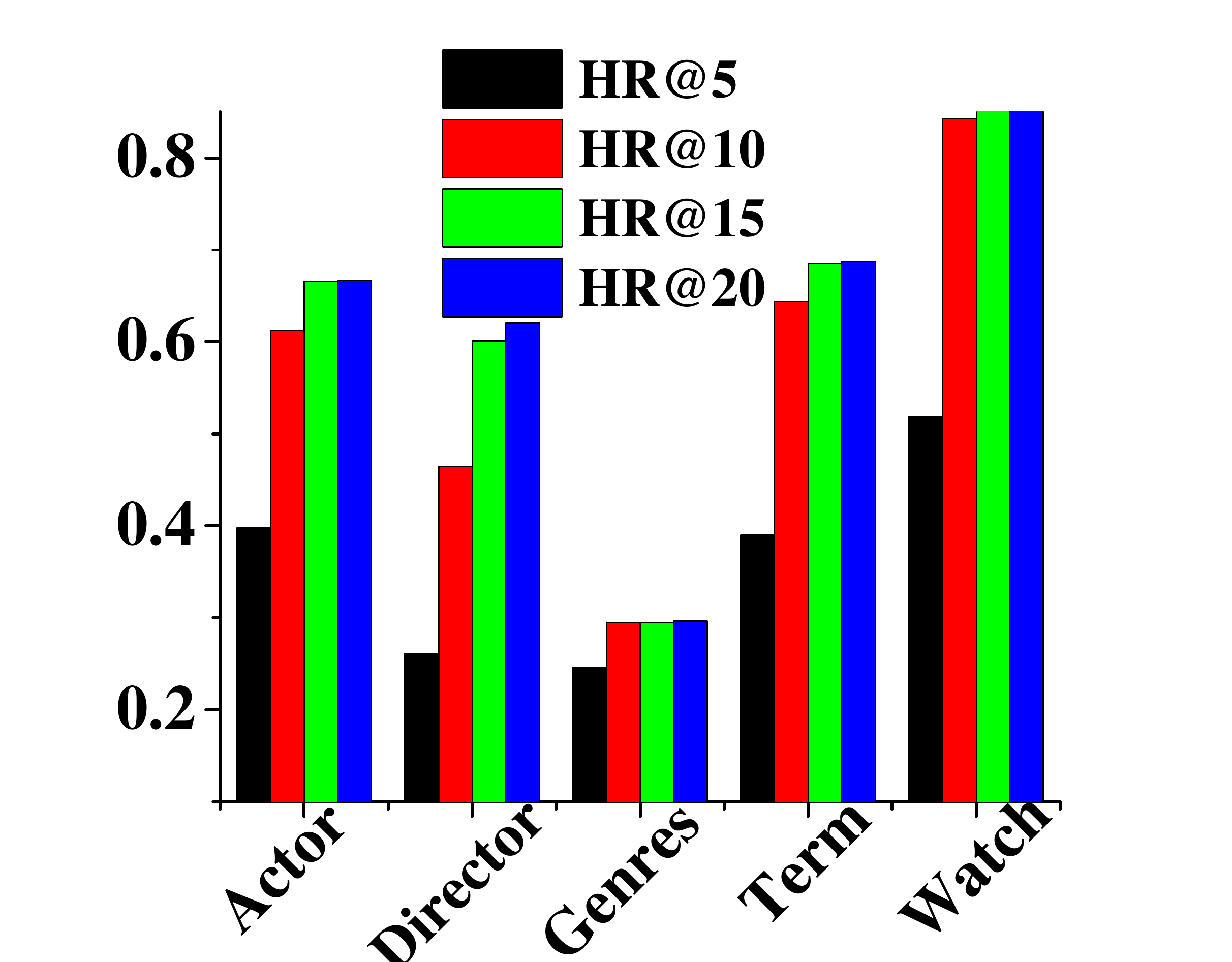}
		\end{minipage}
	}%
	\subfigure[NDCG@$k$ on Movielens.]{
		\begin{minipage}[t]{0.25\linewidth}
			\centering
			\includegraphics[width=1.15\linewidth]{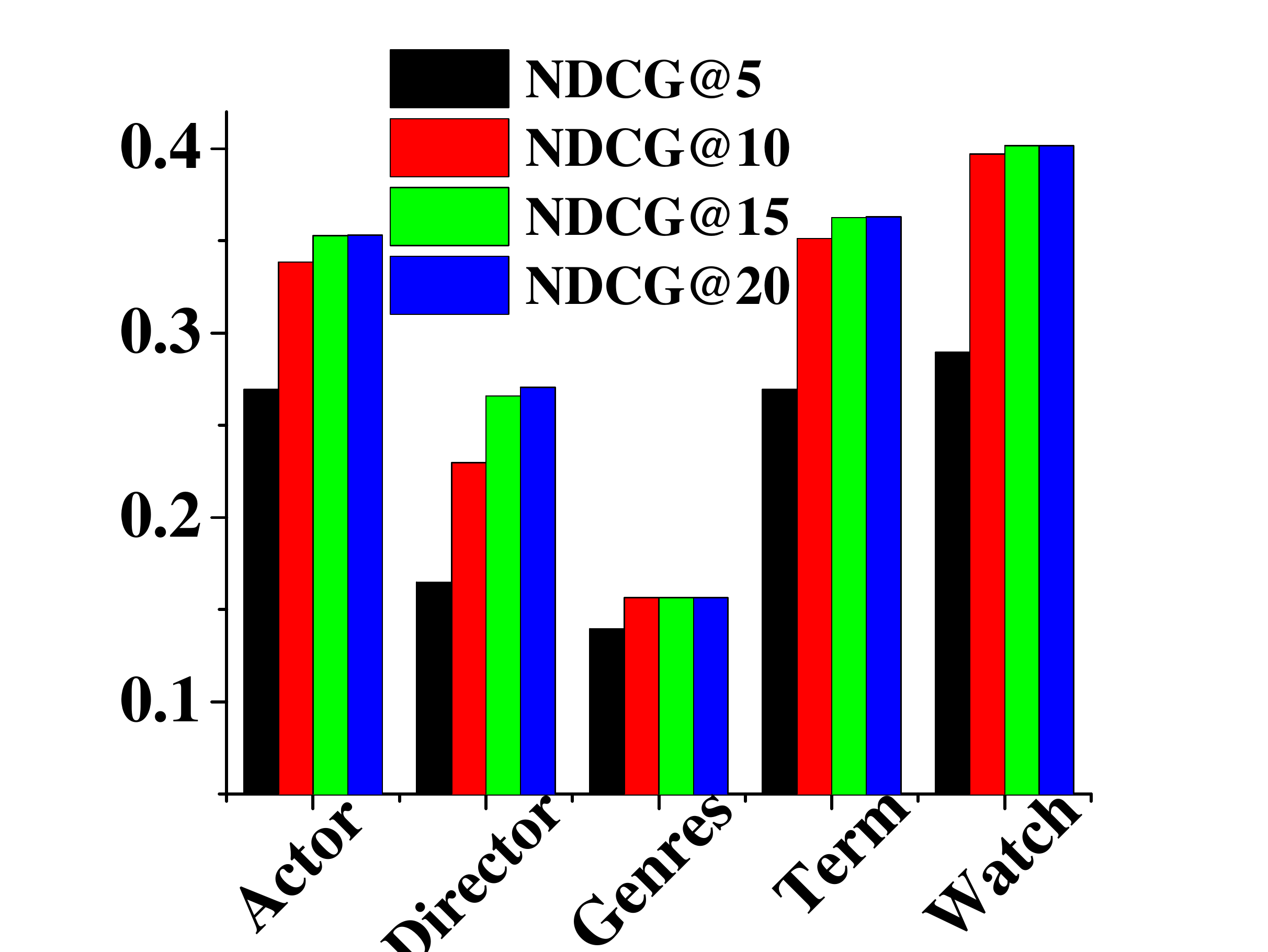}
		\end{minipage}
	}%
	\centering
	\caption{Impacts of different views.}
\end{figure*}

\begin{figure*}
	\centering
	\subfigure[HR@$k$ on Amazon.]{
		\begin{minipage}[t]{0.25\linewidth}
			\centering
			\includegraphics[width=1.15\linewidth]{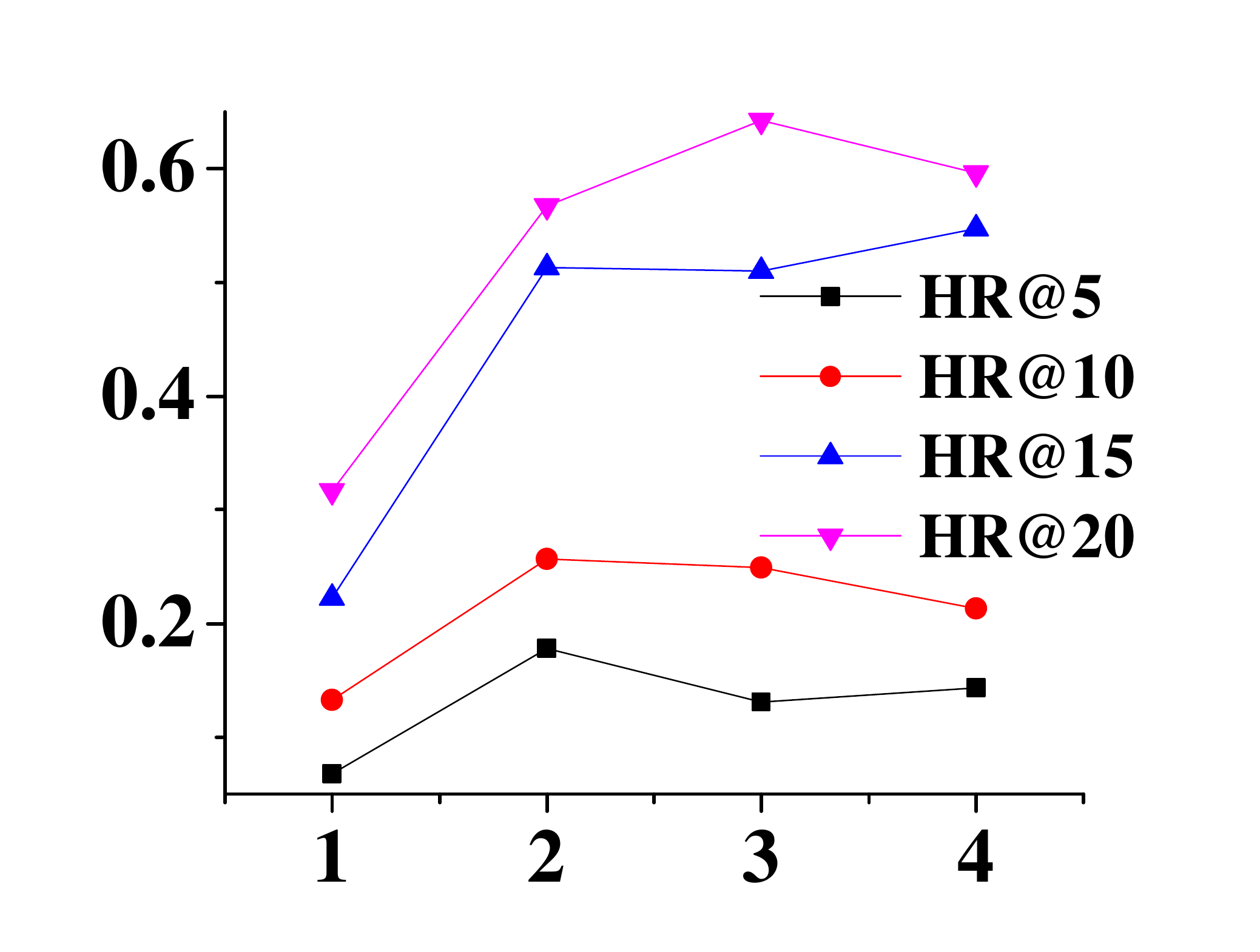}
		\end{minipage}%
	}%
	\subfigure[NDCG@$k$ on Amazon.]{
		\begin{minipage}[t]{0.25\linewidth}
			\centering
			\includegraphics[width=1.15\linewidth]{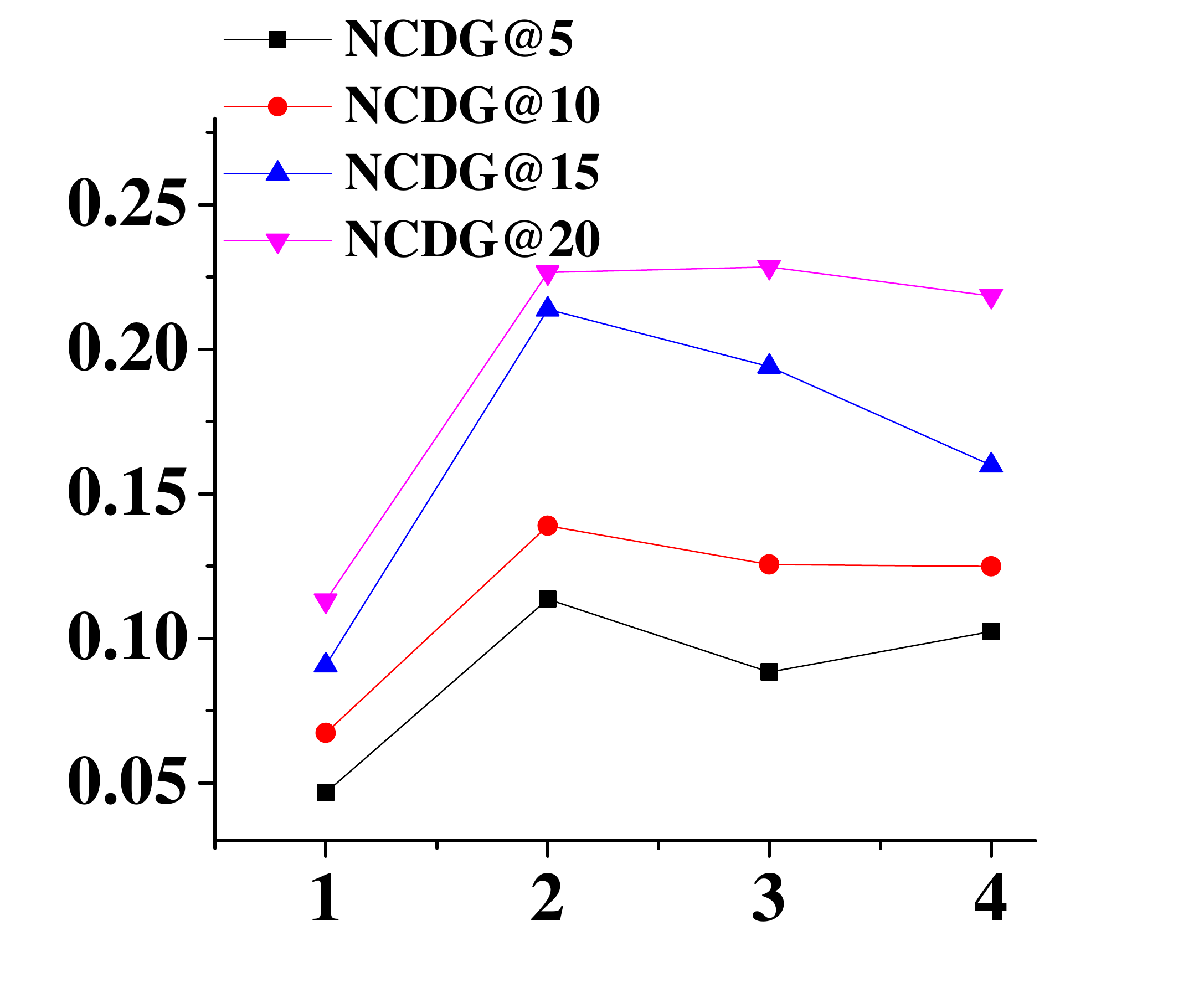}
		\end{minipage}%
	}%
	\subfigure[HR@$k$ on Movielens.]{
		\begin{minipage}[t]{0.25\linewidth}
			\centering
			\includegraphics[width=1.15\linewidth]{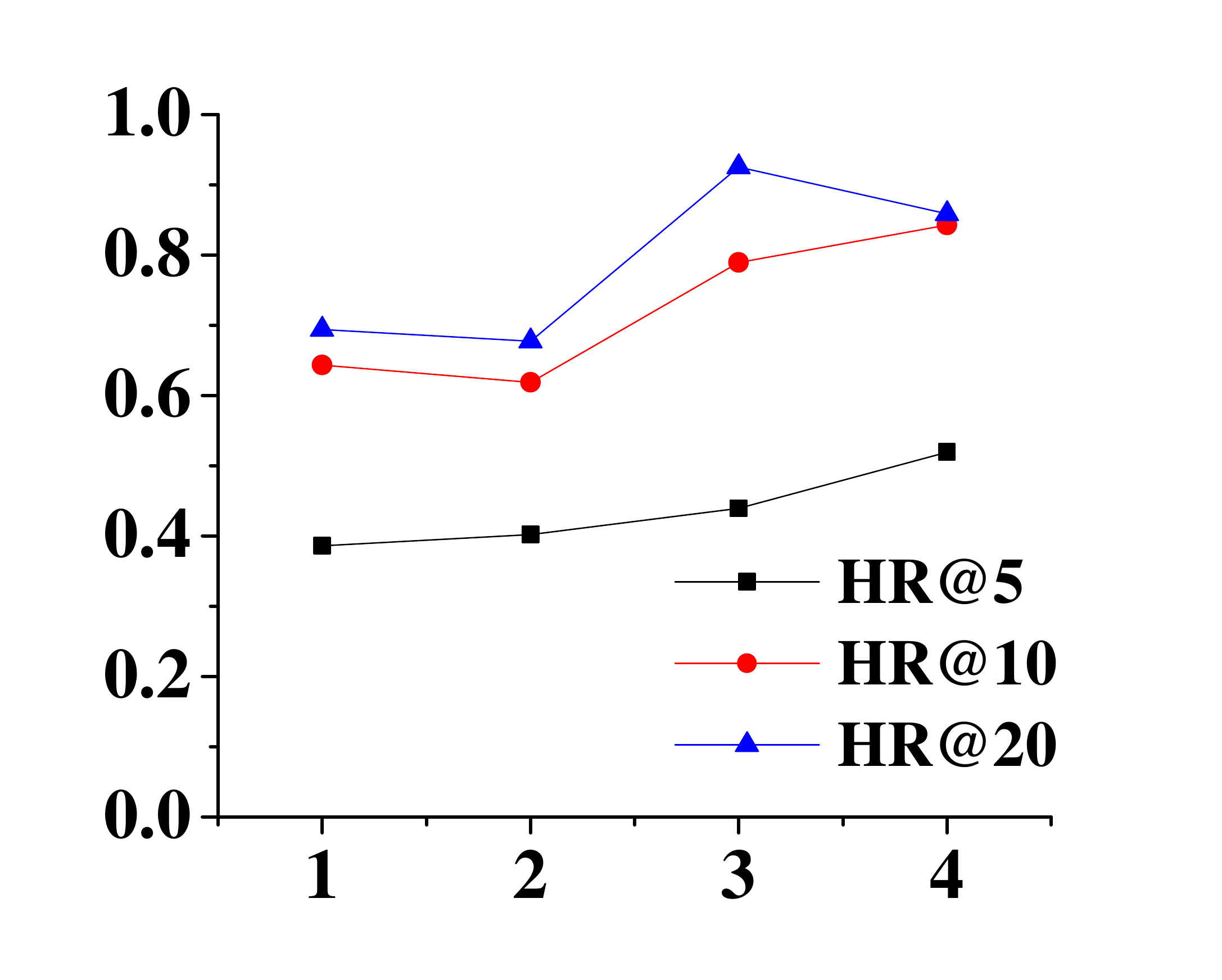}
		\end{minipage}
	}%
	\subfigure[NDCG@$k$ on Movielens.]{
		\begin{minipage}[t]{0.25\linewidth}
			\centering
			\includegraphics[width=1.15\linewidth]{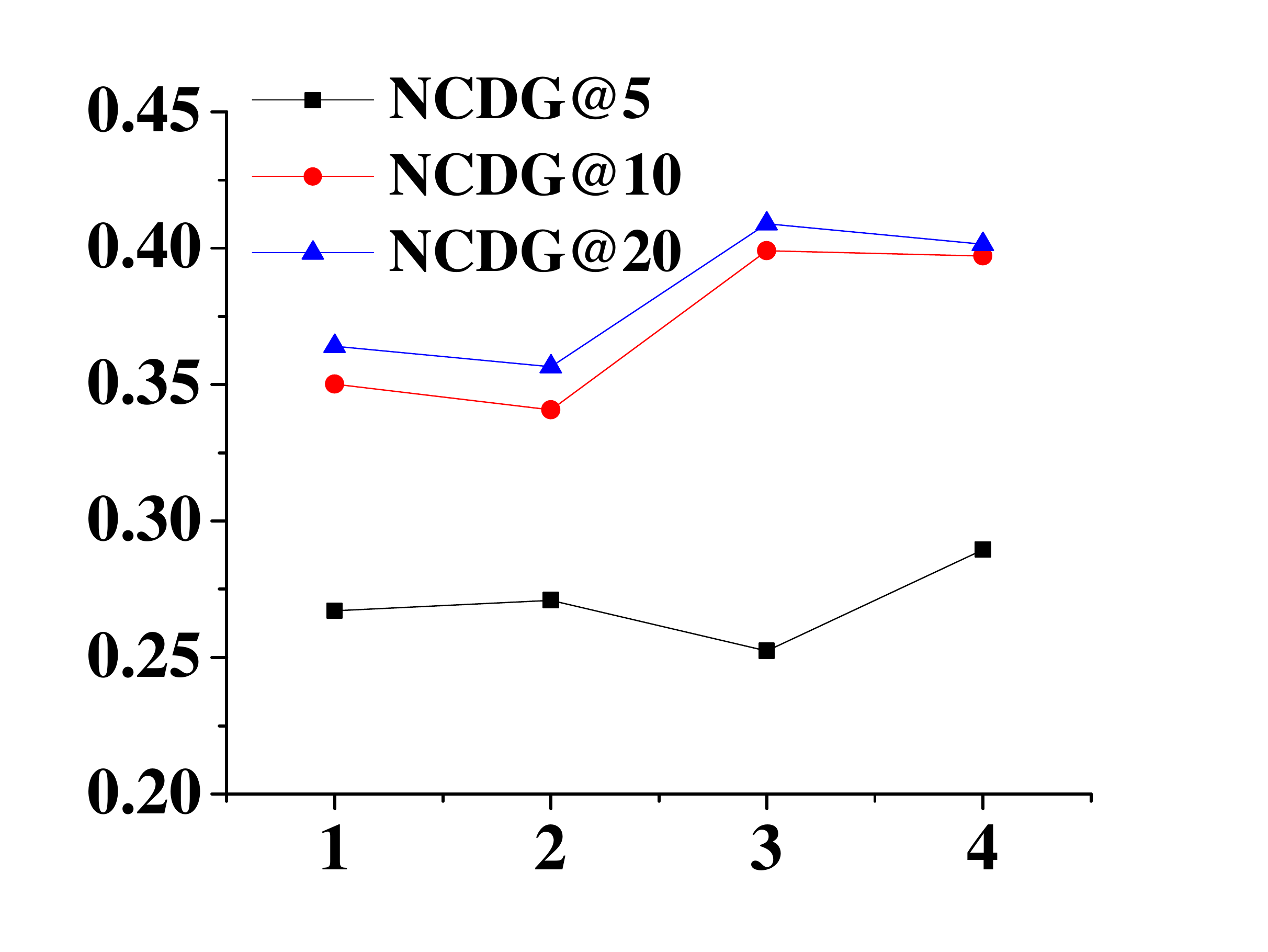}
		\end{minipage}
	}%
	\centering
	\caption{Impacts of different view numbers.}
\end{figure*}

\subsubsection{Dimension of node representations}
\label{sec:decoder}

How to determine the optimal number of embedding dimensions is still an open research problem. Recommendation performances are analyzed when node representation dimension (denoted by $d$ and learned by MDHNE) varies. 

In above figures, the value of HR@k and NDCG@k grows first when the number of dimension continuously increases on two datasets of Amazon and MovieLens datasets. And the performance tends to remain unchanged as the dimension of node representation increases. The reason why such a phenomenon occurs is that this model requires a proper dimension to preserve useful semantic information. If the dimension is excessively large, noisy information may be added, which consequently leads to worse performances and brings extra storage burden. Based on experimental findings above, we set the dimension of node representations as 128 because our proposed MDHNE model needs a suitable dimension for the latent space to encode rich semantic information and too large dimension may introduce additional computational redundancies.

\subsubsection{Length of historical snapshots}
\label{sec:decoder}

We conduct this experiment on Amazon and Movielens datasets to examine how the length of historical snapshots (denoted by $h$ )
affects the performance of MDHNE. We vary the number of historical snapshots from 0 to 4, to demonstrate the effect of varying this parameter. Except for the parameter being tested, all other parameters assume default values.

From the results of MDHNE, we can see the length of historical snapshots affects the HR@k and NDCG@k similarly on different two datasets. The results demonstrate that appropriate length of historical snapshots is essential for training embedding methods to characterize over time. As we can see in Fig.8(a) there is a significant improvement after the number of historical snapshots increases and then becomes stable when the length of historical snapshots reaches 2. Similarly, Fig.8(b) shows that as the number of historical length increases, the performance of our model grows at the beginning and becomes stable when historical length arrives at 3. However, when the length continuously increases, the performance starts to drop slowly. 

\subsubsection{Impacts of Different Views}
\label{sec:decoder}

To analyze the impact of different view on the algorithm performance, we run MDHNE with individual view through setting corresponding meta-path. For example, UIBIU-IBI means that we can learn the "brand" view representations of Users and Items. 

From the results shown in Fig.9(a)(b) on Amazon, one can observe that the "purchase" view (UIU-IUI) can get the better performance than other individual view because this view usually contains the most important information which indicate the purchase history. As is similar with figures aforementioned, experimental results shown in Fig.9(c)(d) on Movielens also demonstrate that "watch" view can get the superior performance over other individual view in MovieLens dataset.

\subsubsection{Number of Views}
\label{sec:decoder}
In addition, analysis is also made about how many views we should consider simultaneously so as to make a balance between better performance and lower computational cost. To further analyze the impact of view numbers, we gradually incorporate add number of views into the proposed model and check the performance changes.

We can observe that generally the performance improves with the incorporation of more views. However, it does not always yield the improvement with more views, and the performance slightly fluctuates. The reason is that some meta-paths may contain noisy or conflict information with existing ones. Moreover, the corresponding performance will stay steadily when number of views continues increasing. In our experiment, two views are taken into account simultaneously on Amazon datasets, and three for Movielens, which are sufficient to demands of most downstream applications. The experiment results also show that our proposed collaborative framework can indeed improve performance by facilitating alignment of different views.

\section{Conclusion and future work}
\label{section7}

The prevalence of heterogeneous information networks in many real-world applications presents new challenges for many learning problems because of its natural heterogeneity and dynamicity. In such networks, interactions among nodes and topological structure tend to evolve gradually. In this paper, we study a novel problem: how to learn embedding representations for nodes in dynamic HIN to further facilitate various mining tasks. Therefore, based on RNN and attention mechanism, we propose a novel framework for incorporating temporal information into HIN embedding methods, denoted as Multi-View Dynamic HIN Embedding (MDHNE), which could efficiently capture evolution patterns of implicit relationships from different views in learning and updating node representations over time. The experiments show that our model can capture temporal patterns on both two real-world datasets and outperform state of the art methods in node classification task and recommendation task. There are several directions for future work: we would like to investigate how the Graph neural network can be applied to dynamic HIN embedding problem. Additionally, we can extend current framework into other types of HIN, including attributed heterogeneous networks, and multi-source heterogeneous information networks.

\section*{Acknowledgments}

	The work was supported by the National Natural Science Foundation of China [grant numbers: 61876138, 61602354]. Any opinions, findings and conclusions expressed here are those of the authors and do not necessarily reflect the views of the funding agencies.

\bibliographystyle{IEEEtran}
\bibliography{refs}

\end{document}